\documentclass[a4paper,11pt]{article}

\usepackage{jheppub} 

\usepackage[T1]{fontenc} 
\usepackage{graphicx}
\usepackage{dcolumn}
\usepackage{bm}
\usepackage{amsmath}
\usepackage{amssymb}
\usepackage{subfigure}%
\usepackage[mathlines]{lineno}
\usepackage{multirow}
\usepackage{amsmath}

\title{\boldmath Measurement of $\theta_{13}$ in Double Chooz using neutron captures on hydrogen with novel background rejection techniques}

\newcommand{\Aachen}{III. Physikalisches Institut, RWTH Aachen University, 52056 Aachen, Germany}
\newcommand{\Alabama}{Department of Physics and Astronomy, University of Alabama, Tuscaloosa, Alabama 35487, USA}
\newcommand{\Argonne}{Argonne National Laboratory, Argonne, Illinois 60439, USA}
\newcommand{\APC}{AstroParticule et Cosmologie, Universit\'{e} Paris Diderot, CNRS/IN2P3, CEA/IRFU, Observatoire de Paris, Sorbonne Paris Cit\'{e}, 75205 Paris Cedex 13, France}
\newcommand{\CBPF}{Centro Brasileiro de Pesquisas F\'{i}sicas, Rio de Janeiro, RJ, 22290-180, Brazil}
\newcommand{\Chicago}{The Enrico Fermi Institute, The University of Chicago, Chicago, Illinois 60637, USA}
\newcommand{\CIEMAT}{Centro de Investigaciones Energ\'{e}ticas, Medioambientales y Tecnol\'{o}gicas, CIEMAT, 28040, Madrid, Spain}
\newcommand{\Columbia}{Columbia University; New York, New York 10027, USA}
\newcommand{\Davis}{University of California, Davis, California 95616, USA}
\newcommand{\Drexel}{Department of Physics, Drexel University, Philadelphia, Pennsylvania 19104, USA}
\newcommand{\Hiroshima}{Hiroshima Institute of Technology, Hiroshima, 731-5193, Japan}
\newcommand{\IIT}{Department of Physics, Illinois Institute of Technology, Chicago, Illinois 60616, USA}
\newcommand{\INR}{Institute of Nuclear Research of the Russian Academy of Sciences, Moscow 117312, Russia}
\newcommand{\CEA}{Commissariat \`{a} l'Energie Atomique et aux Energies Alternatives, Centre de Saclay, IRFU, 91191 Gif-sur-Yvette, France}
\newcommand{\Kansas}{Department of Physics, Kansas State University, Manhattan, Kansas 66506, USA}
\newcommand{\Kitasato}{Department of Physics, Kitasato University, Sagamihara, 252-0373, Japan}
\newcommand{\Kobe}{Department of Physics, Kobe University, Kobe, 657-8501, Japan}
\newcommand{\Kurchatov}{NRC Kurchatov Institute, 123182 Moscow, Russia}
\newcommand{\MIT}{Massachusetts Institute of Technology, Cambridge, Massachusetts 02139, USA}
\newcommand{\MaxPlanck}{Max-Planck-Institut f\"{u}r Kernphysik, 69117 Heidelberg, Germany}
\newcommand{\NotreDame}{University of Notre Dame, Notre Dame, Indiana 46556, USA}
\newcommand{\IPHC}{IPHC, Universit\'{e} de Strasbourg, CNRS/IN2P3, 67037 Strasbourg, France}
\newcommand{\SUBATECH}{SUBATECH, CNRS/IN2P3, Universit\'{e} de Nantes, Ecole des Mines de Nantes, 44307 Nantes, France}
\newcommand{\Tennessee}{Department of Physics and Astronomy, University of Tennessee, Knoxville, Tennessee 37996, USA}
\newcommand{\TohokuUni}{Research Center for Neutrino Science, Tohoku University, Sendai 980-8578, Japan}
\newcommand{\TohokuGakuin}{Tohoku Gakuin University, Sendai, 981-3193, Japan}
\newcommand{\TokyoInst}{Department of Physics, Tokyo Institute of Technology, Tokyo, 152-8551, Japan }
\newcommand{\TokyoMet}{Department of Physics, Tokyo Metropolitan University, Tokyo, 192-0397, Japan}
\newcommand{\Muenchen}{Physik Department, Technische Universit\"{a}t M\"{u}nchen, 85748 Garching, Germany}
\newcommand{\Tubingen}{Kepler Center for Astro and Particle Physics, Universit\"{a}t T\"{u}bingen, 72076 T\"{u}bingen, Germany}
\newcommand{\UFABC}{Universidade Federal do ABC, UFABC, Santo Andr\'{e}, SP, 09210-580, Brazil}
\newcommand{\UNICAMP}{Universidade Estadual de Campinas-UNICAMP, Campinas, SP, 13083-970, Brazil}
\newcommand{\vtech}{Center for Neutrino Physics, Virginia Tech, Blacksburg, Virginia 24061, USA}

\author{Double Chooz Collaboration}
\author[aa]{\\Y.~Abe,}
\author[ac]{S.~Appel,}
\author[e]{T.~Abrah\~{a}o,}
\author[t]{H.~Almazan,}
\author[a]{C.~Alt,}
\author[e]{J.C.~dos Anjos,}
\author[n]{J.C.~Barriere,}
\author[v]{E.~Baussan,}
\author[a]{I.~Bekman,}
% ...

\affiliation[a]{\Aachen}
% ...
\affiliation[e]{\CBPF}
% ...
\affiliation[n]{\CEA}
% ...
\affiliation[t]{\MaxPlanck}
% ...
\affiliation[v]{\IPHC}
% ...
\affiliation[aa]{\TokyoInst}
% ...
\affiliation[ac]{\Muenchen}
% ...

\emailAdd{camil@nevis.columbia.edu, ishitsuka@phys.titech.ac.jp}

\abstract
{The Double Chooz collaboration presents a measurement of the neutrino mixing angle $\theta_{13}$ using reactor $\overline{\nu}_{e}$ observed via the inverse beta decay reaction in which the neutron is captured on hydrogen.
This measurement is based on 462.72 live days data, approximately twice as much data as in the previous such analysis, collected with a detector positioned at an average distance of 1050\,m from two reactor cores.
Several novel techniques have been developed to achieve significant reductions of the backgrounds and systematic uncertainties. 
Accidental coincidences, the dominant background in this analysis, are suppressed by more than an order of magnitude with respect to our previous publication by a multi-variate analysis.
These improvements demonstrate the capability of precise measurement of reactor $\overline{\nu}_{e}$ without gadolinium loading.
Spectral distortions from the $\overline{\nu}_{e}$ reactor flux predictions previously reported with the neutron capture on gadolinium events are confirmed in the independent data sample presented here. 
A value of \mbox{$\sin^{2}2\theta_{13} = 0.095^{+0.038}_{-0.039}$(stat+syst)} is obtained from a fit to the observed event rate as a function of the reactor power, a method insensitive to the energy spectrum shape. 
A simultaneous fit of the hydrogen capture events and of the gadolinium capture events yields a  measurement of \mbox{$\sin^{2}2\theta_{13} = 0.088\pm0.033$(stat+syst)}. 
}

\begin{document}
\maketitle
\flushbottom

\section{Introduction}

In the standard three-flavour framework, the neutrino oscillation probability is described by three mixing angles $\theta_{12}$, $\theta_{23}$, $\theta_{13}$, two independent mass-squared differences, $\Delta m_{21}^{2}$ and $\Delta m_{31}^{2}$, and one CP-violation phase~\cite{ref:PDG}.
The CP-phase and the mass ordering, or hierarchy, of the mass states remain to be determined while all three angles have now been measured.  The angle $\theta_{13}$ has been measured by $\nu_{\mu} \rightarrow \nu_{e}$ appearance in long-baseline accelerator experiments~\cite{ref:MINOS_t13, ref:T2K_t13} and  $\bar\nu_e$ disappearance in short-baseline reactor experiments~\cite{ref:DCIII_nGd, ref:DCII_nH, ref:DCII_RRM, ref:DayaBay, ref:RENO}. In the latter the survival probability, $P$, of $\overline{\nu}_{e}$ with energy $E_{\nu}$\,(MeV)  after traveling a distance of $L$\,(m) can, to a good approximation, be expressed as:
\begin{equation}
P = 1 - \sin^{2}2\theta_{13}\sin^2\left(1.27\,\Delta m_{31}^{2} {\rm (eV^{2})}\,L/E_{\nu}\right).
\label{eq:oscillation}
\end{equation}
The importance of $\theta_{13}$, as well as the other mixing angles, stems from it critically influencing the magnitude of any CP or mass hierarchy effects observable in long-baseline and other experiments.
It is therefore essential for reactor experiments to provide as precise a value of $\theta_{13}$ as possible and cross check themselves to better constrain the inferred value of the CP phase.

Reactor $\overline{\nu}_{e}$'s are observed by a delayed coincidence technique through their inverse $\beta$-decay (IBD) reaction with the free protons in liquid scintillator: \mbox{$\overline{\nu}_e + p \rightarrow e^{+} + n$}.
The positron is observed as the prompt signal arising from its ionisation and subsequent annihilation with an electron. Its energy is related to the neutrino energy by: \mbox{$E_{\rm signal} = E_{\nu} - 0.78\,{\rm MeV}$}.
IBD interactions are tagged via the coincidence between the prompt signal and the delayed signal from the neutron capture on nuclei. 
Current reactor experiments, including Double Chooz~\cite{ref:DCIII_nGd}, which aim to measure $\theta_{13}$ dope their scintillator with gadolinium to benefit from its large neutron capture cross-section resulting in a fast capture time and high energy, about 8\,MeV in total, of its released $\gamma$-rays.
These properties are used to suppress the background from accidental coincidence of natural radioactivity occurring at lower energies, thus justifying the use of gadolinium despite the resulting higher cost and lower light yield due to admixture of gadolinium.
In addition, Double Chooz published the first measurement of $\theta_{13}$ using neutron captures on hydrogen~\cite{ref:DCII_nH}, in which the released $\gamma$-ray carries only 2.2\,MeV, an energy well within the range of natural radioactivity thus leading to sizable background. 

The analysis described in this paper is again based on hydrogen captures (n-H) but it promotes the precision of $\theta_{13}$ measurements to the level achieved with gadolinium captures (n-Gd) through the reduction of background and of systematic uncertainties.
The signal to background ratio was improved from 0.93 to 9.7, more than an order of magnitude, using novel background reduction techniques including accidental background rejection with a neural-network based algorithm.
It uses the same exposure as the recently published $\theta_{13}$ measurement based on n-Gd capture events~\cite{ref:DCIII_nGd} but accumulates about twice the number of events given the 2.2 times larger undoped scintillator volume.
As a consequence of improvements on the systematic uncertainties on the detection efficiency, energy scale and estimate of residual backgrounds, the total uncertainty on the IBD rate measurement was reduced from 3.1\% to 2.3\% of which 1.7\% is associated with the reactor flux prediction.
The value of $\theta_{13}$ is extracted together with the total background rate by fitting the observed IBD rate as a function of the predicted rate, which depends on the reactor power. 
This method is independent of the reactor $\overline{\nu}_{e}$ flux energy distribution, a fact that became important after the observation of unexpected distortions of the reactor flux at about 6\,MeV $\overline{\nu}_{e}$ energy~\cite{ref:DCIII_nGd, ref:DayaBay_Distortion, ref:RENO_Distortion}. 
Double Chooz is particularly well suited for this technique as it is illuminated by only two reactors and variations in reactor power or the turning off of one reactor results in substantial flux variations. 
In addition, during about seven days both reactors were turned off, leading to a very useful direct measurement of the background. 
As a cross check a consistent value of $\theta_{13}$ was also obtained using a fit to the positron energy distribution in spite of the spectrum distortion.

Section~\ref{section:Experiment} describes the experimental setup, section~\ref{section:Reconstruction} the event reconstruction and the determination of the energy scale, section~\ref{section:Selection} the sources of background and the methods to reduce them, section~\ref{section:Background} the residual background estimation, section~\ref{section:Efficiency} the neutron detection efficiency measurement, and section~\ref{section:Analysis} the oscillation analysis. Section~\ref{section:Conclusions} draws the conclusions.
A more detailed description of the Double Chooz detector, simulation Monte Carlo (MC) and calibration procedures can be found in Ref.~\cite{ref:DCIII_nGd}.

\section{Experimental setup}
\label{section:Experiment} 
The far detector (FD) is located at a distance of $\sim$1,050\,m from two reactor cores, each producing  4.25\,GW$_{th}$ thermal power,  of the  \'{E}lectricit\'{e} de France (EDF) Chooz Nuclear Power Plant. It is a liquid scintillator detector made of four concentric cylindrical vessels. 
The innermost volume, named $\nu$ target (NT), is filled with 10.3\,m$^3$ of Gd-loaded liquid scintillator.
NT is surrounded by a 55\,cm thick Gd-free liquid scintillator layer, called $\gamma$ catcher (GC) itself surrounded by a 105\,cm thick non-scintillating mineral oil layer, the Buffer.
The volumes of the GC and Buffer are 22.3\,m$^3$ and 110\,m$^3$, respectively.
The NT and GC vessels are made of transparent acrylic with thickness of 8\,mm and 12\,mm, respectively, while the Buffer volume is surrounded by a steel tank on the inner surface of which are positioned 390 low background 10-inch photomultiplier tubes (PMTs).
They detect scintillation light from energy depositions in the NT and GC. 
Most of the neutron captures on hydrogen occur in the GC, in contrast with the NT where $\sim$85\% occur on gadolinium because of its large capture cross section.
The Buffer works as a shield to $\gamma$-rays from radioactivity of PMTs and surrounding rock.
These inner three regions and PMTs are collectively referred to as the inner detector (ID).
Outside of the ID is the inner veto (IV), a 50\,cm thick liquid scintillator layer viewed by 78 8-inch PMTs, used as a veto to cosmic ray muons and as a shield as well as an active veto to neutrons and $\gamma$-rays from outside the detector.
The detector is surrounded by a 15\,cm thick steel shield to protect it against external $\gamma$-rays. 
A central chimney allows the introduction of the liquids and of calibration sources, which can be deployed vertically down into the NT from a glove box at the detector top.
The calibration sources can be also deployed into the GC using a motor-driven wire attached to the source and guided through a rigid hermetic looped tube (GT).
The loop passes vertically near the GC boundaries with the NT and Buffer down to the centre of the detector.

Signal waveforms from all ID and IV PMTs are digitized at 500\,MHz by 8-bit flash-ADC electronics~\cite{ref:FADC_NL}.
The trigger threshold is set at 350\,keV, well below the  1.02\,MeV minimum energy of $\overline\nu_e$ signals.
 
An outer veto (OV) consisting of two orthogonal layers of 320\,cm $\times$  5\,cm $\times$ 1\,cm scintillator strips covers an area of 13\,m $\times$ 7\,m on top of the detector except for a gap around the chimney covered by two smaller layers mounted above the chimney. Of the data presented here, 27.6\% were taken with the full OV, 56.7\% with only the bottom layers and 15.7\% with no OV.

Neutron and gamma sources have been used to calibrate the energy scale and to evaluate the detection systematics, including the neutron detection efficiency and the fraction of hydrogen in the liquid scintillator.
Laser and LED systems are used to measure the time offset of each PMT channel and its gain.

Double Chooz has developed a detector simulation based on Geant4~\cite{ref:Geant4} with custom models for neutron thermalisation, scintillation processes, photocathode optical surface, collection efficiency of PMT and readout system simulations based on measurements.

The data used here include periods in which both reactors, only one reactor or no reactor were in operation.
The $\overline{\nu}_{e}$ flux is calculated by the same way as in Ref.~\cite{ref:DCIII_nGd} using locations and initial burn-up of each fuel rod assembly and instantaneous thermal power of each reactor core provided by EDF.
Reference $\overline\nu_{e}$ spectra for three of the four isotopes producing the most fissions, $^{235}$U,  $^{239}$Pu and $^{241}$Pu, are derived from measurements of their $\beta$ spectrum at ILL~\cite{ref:SchreckU5,ref:SchreckU5Pu9,ref:SchreckPu9Pu1}. 
A measurement~\cite{ref:Haag} of the $\beta$ spectrum from $^{238}$U, the fourth most prolific isotope, is used in this analysis.
Evolution of each fractional fission rate and associated errors are evaluated using a full reactor core model and assembly simulations developed with the MURE simulation package~\cite{ref:MURE, ref:MURE-NEA}.
Benchmarks tests have been performed with other codes~\cite{ref:FluxCheck} in order to validate the simulations. 
By using as normalisation the $\overline{\nu}_{e}$ rate measurement of Bugey4~\cite{ref:Bugey4} located at a distance of 15\,m from its reactor, after corrections for the different fuel composition in the two experiments, the systematic uncertainty in the $\overline\nu_e$ prediction was reduced to 1.7\% of which 1.4\% is associated with the Bugey4 measurement.

\section{Vertex position reconstruction and energy scale}
\label{section:Reconstruction}
The same vertex position reconstruction algorithm and energy scale as in the n-Gd analysis~\cite{ref:DCIII_nGd} are used in the analysis described in this paper, while the systematic uncertainty on the energy scale is newly estimated to account for differences between the GC and the NT.

The charge and timing of signals in each PMT are extracted from the  waveform digitized by the flash-ADCs. 
The integrated signal charge is defined as the sum of ADC counts over the 112\,ns integration time window after baseline subtraction.
The integrated signal charge is then converted into the number of photoelectrons (PE) based on the gain calibration in which non-linearity of the gain introduced by the digitisation is taken into account.
The vertex position of each event is reconstructed using a maximum likelihood algorithm based on the number of PE and time recorded by each PMT, assuming the event to be point-like. 
A goodness of fit parameter, F$_{\rm V}$, is used to evaluate the consistency of the fit with the point-like behaviour expected from electrons and positrons of a few MeV. 

The absolute energy scale is determined by deploying, in the centre of the detector, a $^{252}$Cf source emitting neutrons and observing the 2.2\,MeV peak resulting from their capture by the scintillator hydrogen.
The energy scale is found to be 186.2 and 186.6 p.e./MeV for the data and MC respectively. 
The visible energy, $E_{\rm vis}$, of every event is then obtained by correcting its total number of photoelectrons for uniformity, time stability and charge non-linearity as discussed below.
Reconstruction and the correction of the visible energy in the MC simulation follow the same procedures as in the data, although the stability correction is applied only to the data and the charge non-linearity correction is applied only to the MC.
By definition, $E_{\rm vis}$ represents the single-$\gamma$ energy scale which is relevant for the delayed signal.

The non-uniformity of the energy response over the detector is corrected for using n-H captures collected from muon spallation. 
They are split into two independent samples interleaved in time to avoid time variation effects. 
Two independent neutron capture samples were also simulated by the MC. 
Using the first samples, the uniformity corrections are obtained separately for the data and MC by comparing the energy response at each position to that at the centre.
After applying these corrections, a uniformity correction uncertainty of 0.25\% is obtained from the RMS of the remaining difference between the second data and MC samples.

The time variation of the mean gain in the data is corrected using the spallation n-H capture peak.
The correction is applied with a linear dependence on energy determined using values of the hydrogen and Gd (8\,MeV) spallation neutron capture peaks and of the 8\,MeV $\alpha$ from $^{212}$Po decays originating from the $^{212}$Bi-$^{212}$Po decay chain, which appears at $\sim$1\,MeV due to quenching.
A stability systematic uncertainty of 0.34\% is estimated based on the $\alpha$, n-H IBD captures and n-Gd spallation captures residual variations, weighted over the IBD prompt energy spectrum. 
It was 0.50\% in n-Gd analysis~\cite{ref:DCIII_nGd} using n-Gd IBD captures with poorer statistics.

\begin{figure}
\begin{center}
\includegraphics[width=100mm]{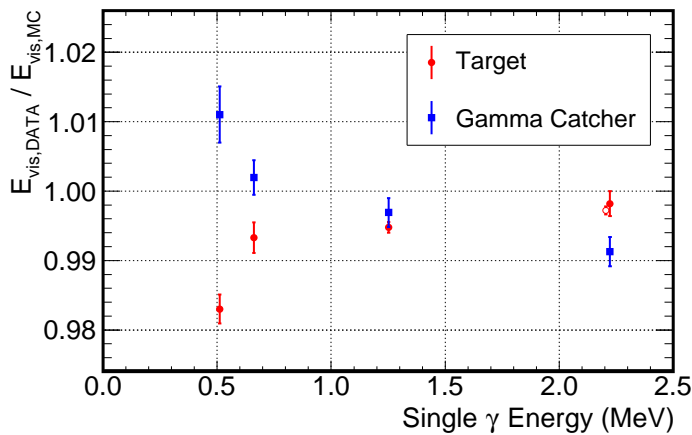}
\caption{Remaining discrepancy of the energy scales between the data and MC after all corrections including charge non-linearity are applied. Points shows the ratio between the data and MC of the visible energy of, from left to right, $^{68}$Ge, $^{137}$Cs, $^{60}$Co and$^{252}$Cf sources plotted as a function of the averaged single $\gamma$ energy. The sources were deployed at the centre of the ND (red circles) and around the middle of the GC layer (blue squares). The points at around 2.2\,MeV refer to the n-Gd captures in the NT (open red circle), n-H captures in the NT (solid red circle) and n-H captures in the GC (solid blue square) with neutrons emitted from the $^{252}$Cf source.}
\label{fig:LNL}
\end{center}
\end{figure}

Non-linearity arises from both charge non-linearity (due to readout and charge integrating effects) and scintillator light non-linearity. 
The first is corrected for by comparing the detector response to the 2.2\,MeV $\gamma$-rays from n-H captures and to the 8\,MeV release of n-Gd captures. 
As the average energy of $\gamma$-rays emitted in n-Gd captures is about 2.2\,MeV, an energy almost the same to that of the $\gamma$-ray from n-H capture, the discrepancy of the energy response between the data and MC can be understood to be due to charge integration rather than to scintillator light yield.
After the charge non-linearity is corrected, the residual non-linearity is attributed to the scintillator light non-linearity. 
It is evaluated by comparing the measured energy of $\gamma$'s of known energy from various sources in the data and MC. 
As shown in Figure~\ref{fig:LNL}, it differs between the NT and GC as they are filled with different scintillators. 
Unlike the previous publication using neutrons captures on gadolinium occurring in the NT, scintillator light non-linearity is not corrected for in the n-H sample.
Instead, in the Rate+Shape fit using the energy spectrum of the prompt positron signal (Section~\ref{section:RS}), the uncertainty on the scintillator light non-linearity is taken to cover the possible variation evaluated by the source calibration data and is left to be determined within the fit to the energy spectrum.
We confirmed the output parameters for the non-linearity correction obtained from a R+S fit to the n-Gd sample with this new approach are consistent with the correction we applied in the previous publication.
The systematic uncertainty on the energy scale at 1.0\,MeV (lower cut of the prompt energy window) is evaluated to be 1.0\%, which results in the IBD rate uncertainty of 0.1\% caused by the prompt energy cut. 

\section{Neutrino Selection}
\label{section:Selection}
An IBD interaction is characterized by the prompt positron energy deposit followed within a few hundred $\mu$s by the delayed energy deposit of the $\gamma$-ray(s) released by neutron capture, in this case by hydrogen. 
Two types of backgrounds, accidental coincidence of two uncorrelated signals and two consecutive correlated signals, can simulate IBD interactions and thus affect the measurement of $\overline{\nu}_{e}$ disappearance.
They are reduced by the coincidence condition and other dedicated vetoes for each background source described in this section. 
Table~\ref{table:MCIneffCF} summarizes them as well as the backgrounds they target.
Vetoes in Table~\ref{table:MCIneffCF}, except for the coincidence condition, are applied only to the data as the muons and light noise are not simulated in the IBD signal MC.
Instead, corrections for the resulting veto inefficiencies are applied to the MC.
Efficiencies of the IBD signal and the systematic uncertainties are evaluated from the data and listed in Table~\ref{table:MCIneffCF}. 

The final IBD candidates used in the neutrino oscillation analysis were selected by the combination of vetoes summarized in Table~\ref{table:MCIneffCF} and explained below. 
These vetoes are based on the response from different detectors (ID, IV and OV) and hence complementary without correlations in the rejected events.

The prompt energy window is set to $1.0 \leq E_{\rm vis} \leq 20.0\,{\rm MeV}$.
One of the two $\gamma$-rays from the annihilation of a positron produced by an IBD interaction in the buffer volume often enters the GC.
In our gadolinium analysis the lower cut was 0.5\,MeV as these buffer events would not be selected as IBD candidates as it is unlikely for a neutron produced in the buffer to travel as far as the NT to be captured on gadolinium. 
In this analysis however one of the two $\gamma$-rays from buffer could be identified as a prompt signal peaking at 0.5\,MeV if it is followed by a delayed signal due to the neutron capture on hydrogen in the GC or the buffer.
A cut at 0.5\,MeV would include only partially this $\gamma$ signal. 
Since reducing the cut would run into our trigger threshold of 0.35\,MeV, it was decided instead to exclude these $\gamma$'s by increasing the lower cut to 1.0\,MeV.
The prompt signal from reactor $\overline{\nu}_{e}$ extends to around 8\,MeV while the energy window is extended up to 20\,MeV to better constrain the background due to cosmogenic isotopes and fast neutrons (FN) using their different energy spectrum shapes.

The live time of the detector is calculated to be 462.72 live days after the muon veto and OV veto are applied.

\begin{table}[ht]   
  \begin{center}
    \begin{tabular}{| c|c|c|c |}
      \hline
      Cut & Target of cut & MC corr. & Uncer.(\%) \\
      \hline
       Muon veto & muons and their cosmogenic isotopes & 0.9400 & $<$ 0.01 \\
       LN cut & spontaneous light emission by PMT & 0.9994 & $<$ 0.01 \\
       Coincidence condition$^{*}$ & \multirow{2}{*}{single, accidental coincidence} & \multirow{2}{*}{1.0000} & \multirow{2}{*}{0.220} \\
       (ANN cut) & & & \\
       Multiplicity cut & multiple $n$ scattering and captures & 0.9788 & $<$ 0.01 \\
       $\rm{F_{V}}$ veto & stopped $\mu$, spontaneous light emission & 0.9995 & 0.015 \\
       Li veto & cosmogenic isotopes ($^{9}$Li, $^{8}$He, $^{12}$B) & 0.9949 & 0.012 \\
       OV veto & fast $n$, stopped $\mu$ & 0.9994 & 0.056 \\
       IV veto & fast $n$, stopped $\mu$, $\gamma$ scattering & 1.0000 & 0.169 \\
       MPS veto & fast $n$ & 1.0000 & 0.100 \\
      \hline
   \end{tabular}
  \end{center}
  \caption{Summary of cuts to select n-H IBD candidates and the correction factors applied to the MC to account for the inefficiencies introduced by each cut. *Unlike the others, coincidence condition was applied to both the data and MC, with the same IBD efficiency on both, resulting in a correction factor of unity with the quoted uncertainty (see Section~\ref{section:Efficiency}).}
  \label{table:MCIneffCF}
\end{table}

{\bf Muon veto}: 
Defining a muon as an energy deposit in the ID greater than 20\,MeV or in the IV greater than 16\,MeV\footnote{MeV-equivalent energy scale reconstructed from the integrated charge in the IV.}, no energy deposit is allowed to follow a muon by less than 1.25\,ms.
20\,MeV and 16\,MeV correspond to approximately 11\,cm and 9\,cm path length by a MIP in the ID and IV, respectively.
Inefficiency due to the muon veto is computed to be 6.0\% with negligible errors by measuring the live time after the muon veto is applied.

{\bf  Light noise (LN) cut}: 
Random light releases by PMT bases are eliminated by the same cuts as in the n-Gd analysis~\cite{ref:DCIII_nGd}.
They reject energy depositions concentrated in a few PMTs and spread out in time.
This results in an inefficiency of \mbox{(0.0604 $\pm$0.0012)\,\%}. 
 
\begin{figure}
\includegraphics[width=70mm]{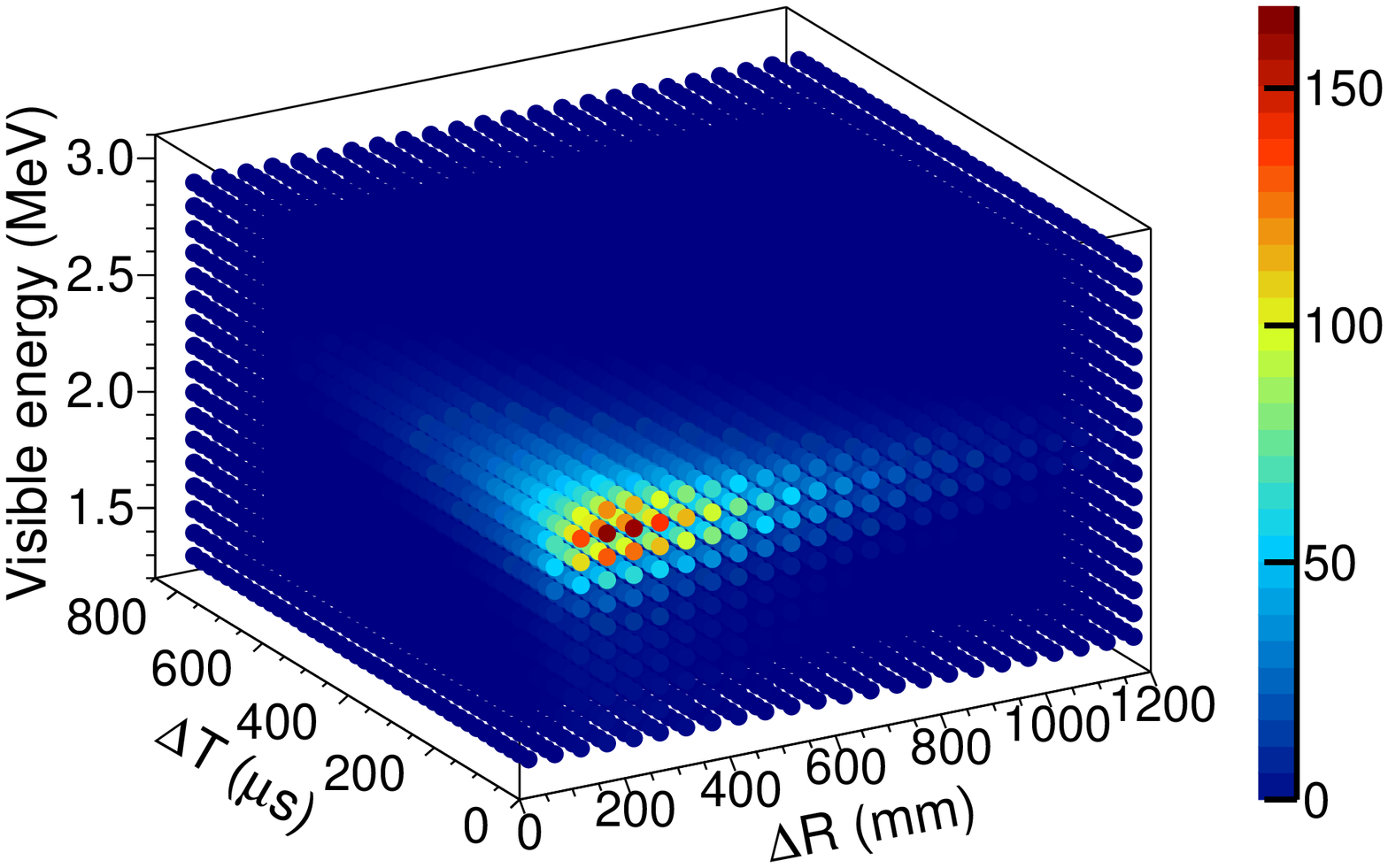}
\includegraphics[width=70mm]{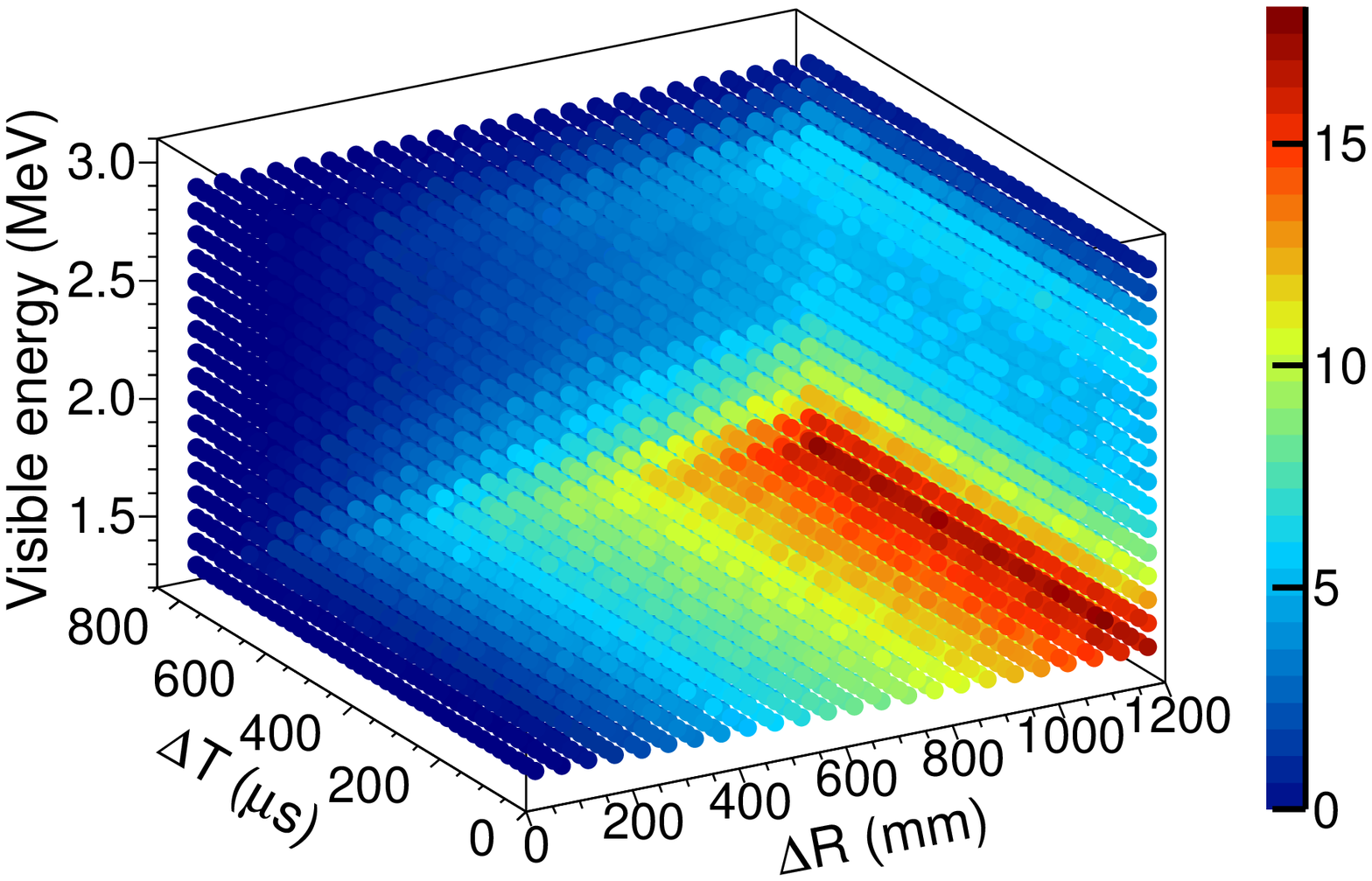}
\caption{Three-dimensional distributions of  $E_{\text{delayed}}$, $\Delta T$ and  $\Delta R$ for MC signal events (left plot) and random associations of off-time events (right plot) showing the different patterns of signal and random association events. In the right plot the $\Delta T$ shown is after subtraction of the time offset used in the random associations.}
\label{fig:ANA3D}
\end{figure}

 {\bf ANN cut}:
Random associations of two energy deposits can simulate IBD events.
This uncorrelated background is much more frequent in hydrogen capture than in gadolinium capture events as the low energy (2.2\,MeV) of the capture $\gamma$ is in an energy range highly populated by ambient and PMT radioactivity. 
In our previous analysis, to reduce it, sequential cuts on the energy of delayed signal, $ E_{\rm delayed}$, and on the time and spatial differences between the prompt and delayed energies,  $\Delta$T and $\Delta$R, were used. 
These differences are illustrated as three-dimensional plots of  $ E_{\rm delayed}$ vs  $\Delta$T vs $\Delta$R in Figure~\ref{fig:ANA3D} for MC signal events (left plot) and for accidental associations of events in which the delayed time window is shifted by a time offset of more than 1\,s (right plot), referred to as off-time. 

\begin{figure}
\includegraphics[width=49mm]{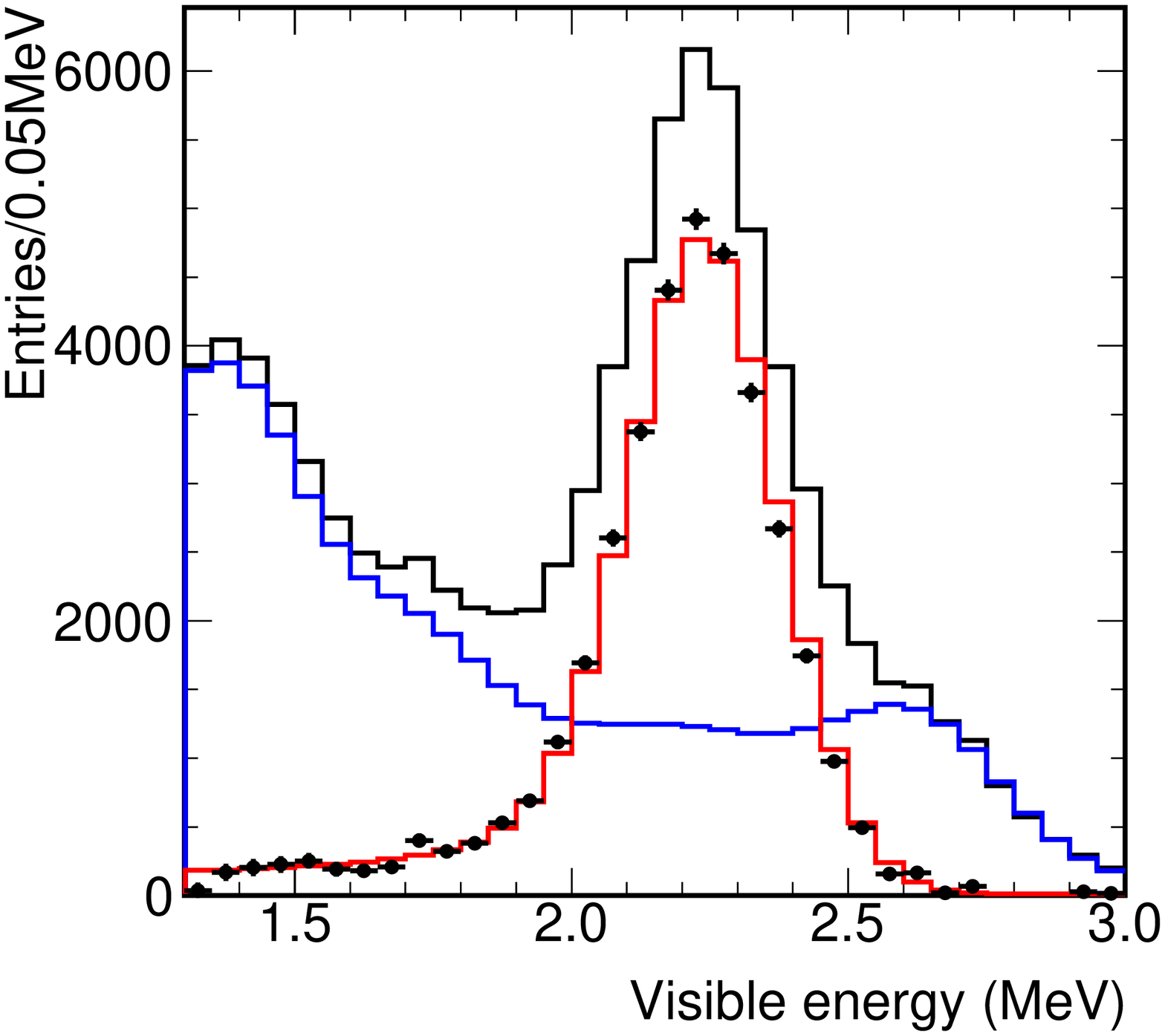}
\includegraphics[width=49mm]{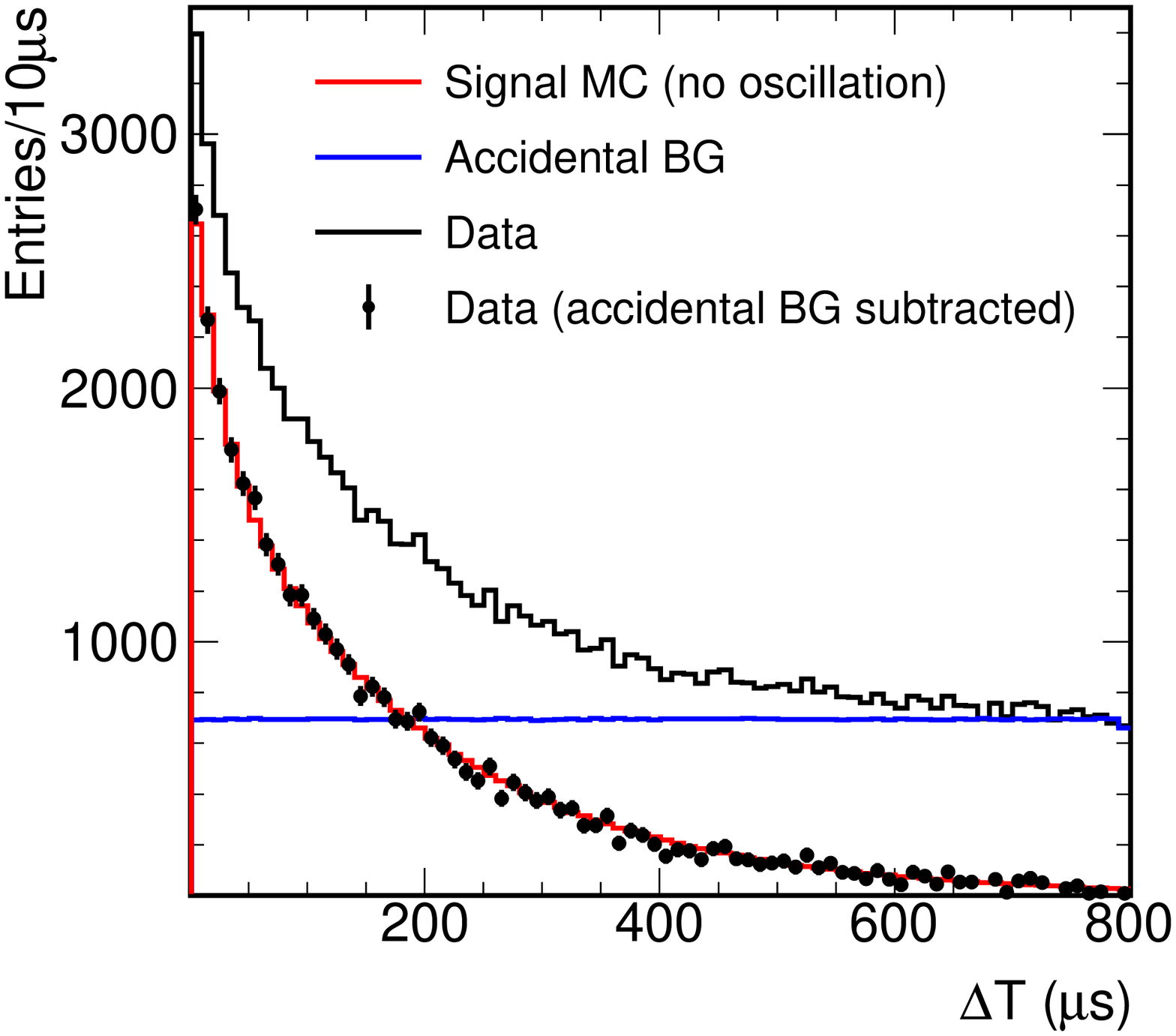}
\includegraphics[width=49mm]{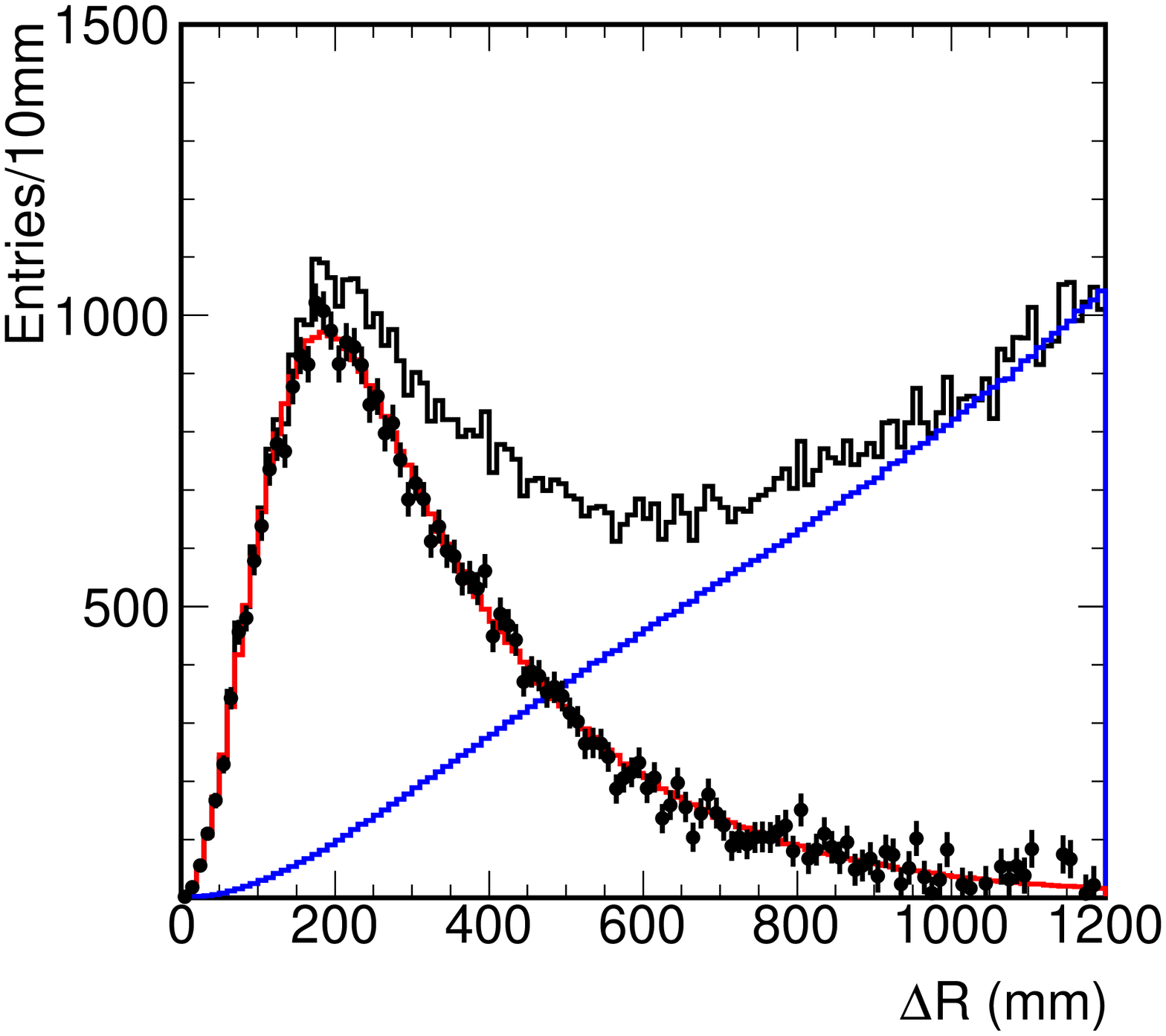}
\caption{Input variables to ANN: $E_{\text{delayed}}$ (left), $\Delta T$ (centre) and  $\Delta R$ (right) for the IBD signal MC (red), accidental background from off-time coincidences (blue) and the on-time data before (black histogram) and after (points) subtraction of the accidental background.}
\label{fig:ANN1D}
\end{figure}

To benefit from these notable differences between the signal and random background distributions a multivariate analysis based on an artificial neural network (ANN) was implemented. 
Three variables, $\Delta$R, $\Delta$T and  $E_{\rm delayed}$ were used as the input to ANN after confirmation of the agreement between the data and MC simulation as shown in Figure~\ref{fig:ANN1D}.
The ANN used was the MLP (Multi-Layer Perceptron) network with Back Propagation from the TMVA package in ROOT~\cite{ref:TMVA}. 
The network structure included an input layer with four nodes (three input variables $+1$ bias node, whose value is constant  and the weight is adjusted during the training to optimize the output), a single hidden layer with 9 nodes and a single output parameter.
A hyperbolic tangent was used as the neuron activation function and resulted in a continuous output in the range $-1.2$ to $+1.2$. 
The neural network was trained using an IBD MC sample for the signal and a sample obtained from off-time coincidences for the accidental background. 
After training, different samples were used for testing the neural network.

\begin{figure}
\includegraphics[width=75mm]{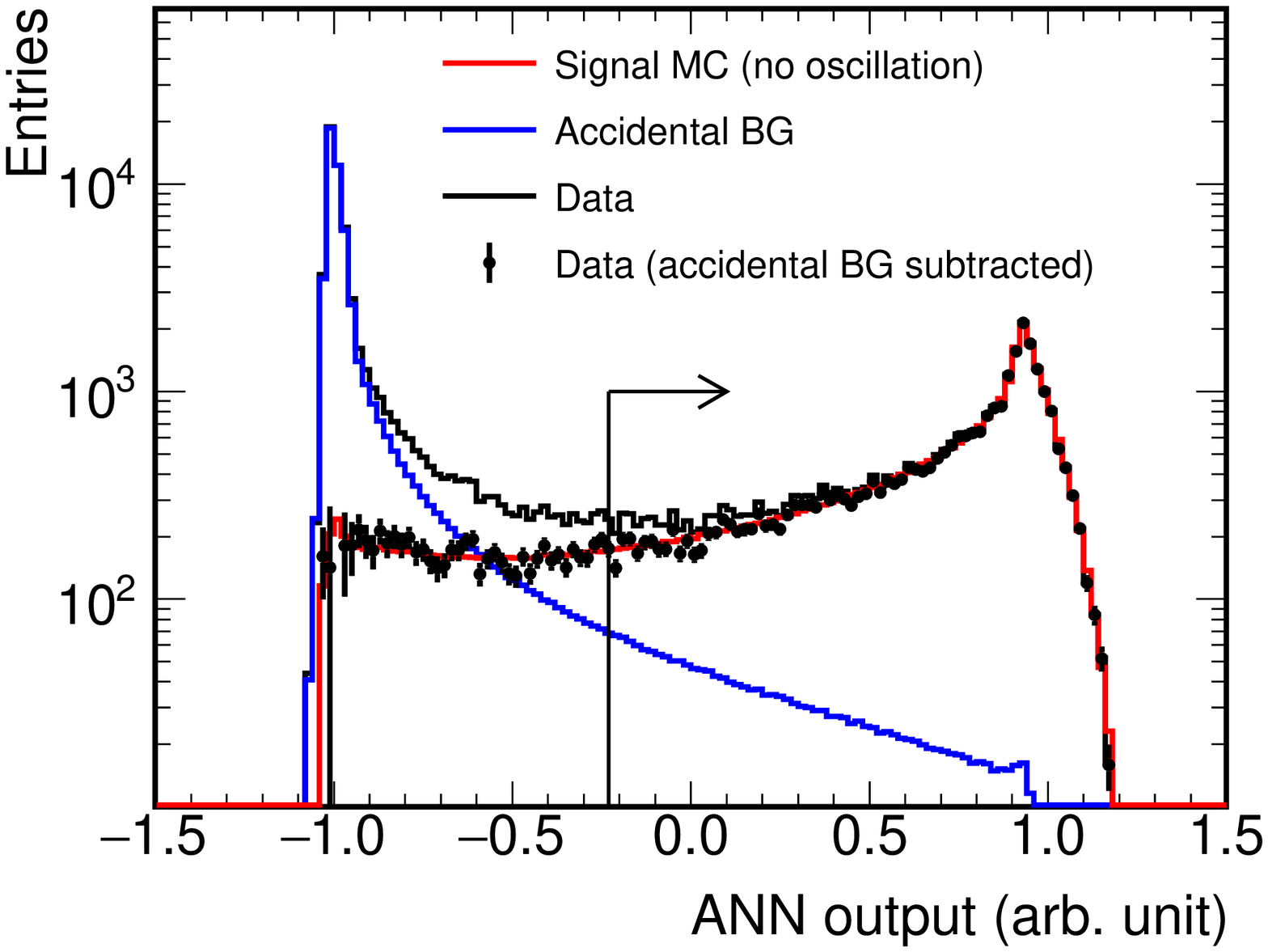}
\includegraphics[width=75mm]{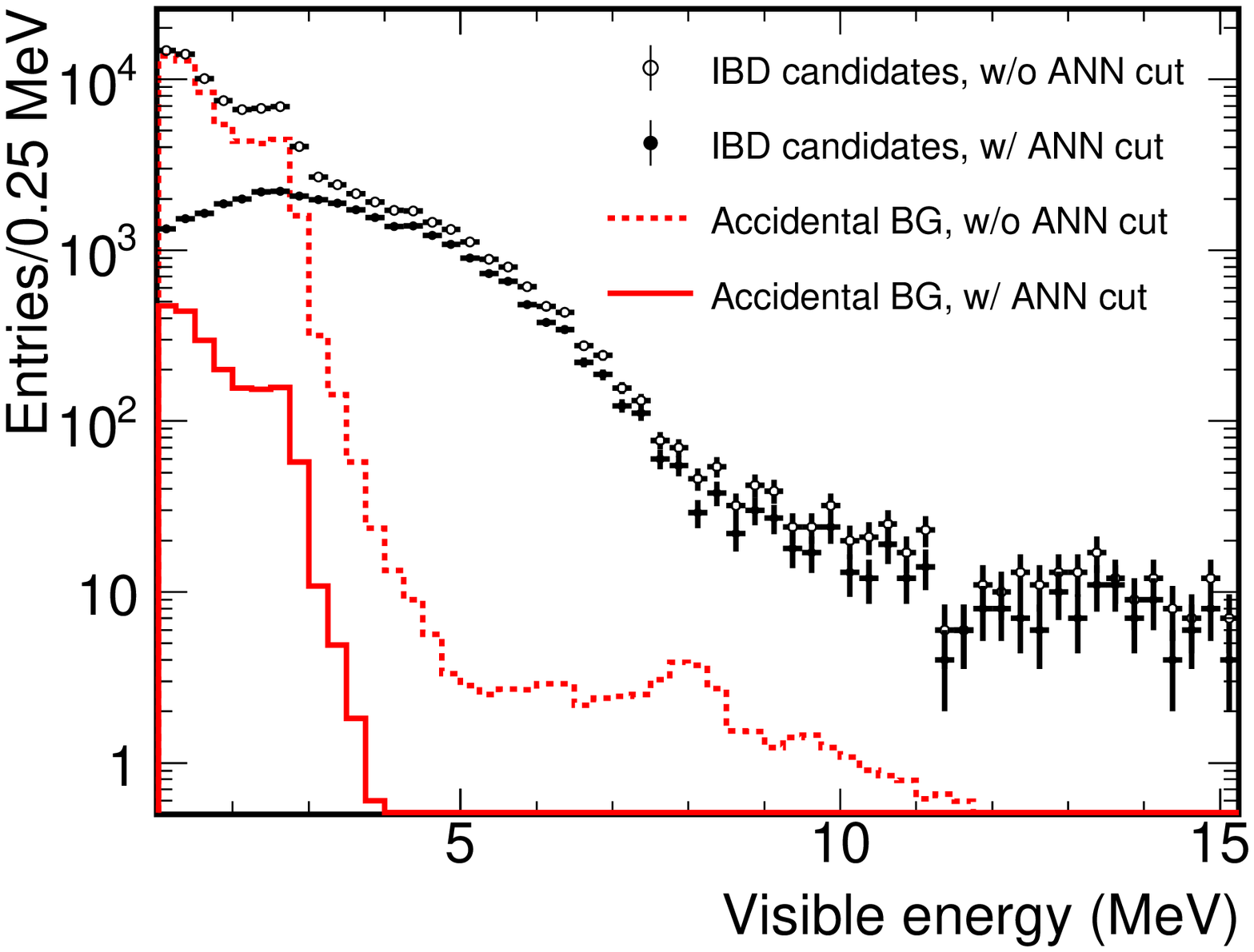}
\caption{\label{fig:ANNeffect}Left: ANN classifier output for the on-time data (black histogram), off-time data (blue), on-time minus off-time data (black points) and signal MC (red). Right: The prompt energy distributions of IBD candidates (black) and accidental events (red) before and after the application of the ANN cut indicated by the arrow in the left-hand plot.}
\end{figure}

The ANN output is shown in Figure~\ref{fig:ANNeffect} (left) for on-time and off-time delayed coincidence data.
The difference between off-time and on-time data is seen to agree very well with the MC signal, also shown in the figure. 
A cut of ANN $\geq -0.23$  was applied, together with \mbox{1.3\,MeV $\leq E_{delayed} \leq$ 3.0\,MeV}, \mbox{0.50\,$\mu$s $\leq \Delta$T $\leq$ 800\,$\mu$s}, \mbox{$\Delta$R $\leq$ 1200\,mm}.
By replacing sequential cuts used in our previous hydrogen capture publication~\cite{ref:DCII_nH} with ANN, the signal to accidental background ratio is improved by more than a factor of seven while the IBD efficiency only decreased by $\sim$6\%. 
The prompt spectrum of IBD candidates (black) and the accidental background (red) are shown in Figure~\ref{fig:ANNeffect} (right) before and after the ANN cut, clearly demonstrating its effectiveness.
Its application greatly reduces the accidental background and allows the IBD signal to dominate the distribution.
The accidental background is further reduced using the IV cut described below.

\vspace{1mm}
Some of the major backgrounds are caused by the interactions of cosmic muons in or close to the detector, resulting in the production of neutrons and isotopes (cosmogenic).
Muon generated events are therefore vetoed as follows:

{\bf Multiplicity cut}: In order to reject cosmogenic background events due to multiple neutron captures, no energy deposits other than the prompt and delayed candidates were allowed from 800\,$\mu$s preceding the prompt to 900\,$\mu$s following it.
Random associations of an IBD event with an additional energy deposit results in an IBD inefficiency of 2.12\% calculated from the 13.2\,s$^{-1}$ singles rate measured in the detector after LN cut and muon veto are applied.

{\boldmath $F_{\rm V}$} {\bf veto}: 
Muons can enter the detector through the chimney, undetected by the OV and IV and then stop in the ID (stopping muons, SM). 
In a delayed coincidence with their decay electron they can simulate IBD events. 
The large $F_{\rm V}$ of the Michel electron being confined in the chimney or of the remaining light noise after the LN cut indicate inconsistency of these backgrounds with the point-like hypothesis in the vertex reconstruction (Section~\ref{section:Reconstruction}). 
The IBD candidates for which the delayed signal satisfy $E_{\rm delayed} \geq 0.276 \times \exp(F_{\rm V}/2.01)$ are selected.
This introduces an IBD inefficiency of $(0.046 \pm 0.015)\,\%$ estimated from the number of IBD candidates rejected by the $F_{\rm V}$ veto, after subtracting SM and LN components. 

{\bf Li veto}: Muons entering the detector and undergoing spallation interactions, can produce $^9$Li and $^8$He (collectively referred to as Li) which then $\beta$ decay with the subsequent emission of a neutron, perfectly simulating an IBD event.
This is often accompanied by additional neutrons depositing a few MeV within 1\,ms of the muon.
The long lifetimes of $^9$Li and $^8$He (257\,ms  and 172\,ms, respectively) prohibit their rejection by vetoing on an entering muon.
Instead, a likelihood based on the distance between the event vertex position and a muon track and on the number of neutron candidates following the muon within 1\,ms is used to identify the cosmogenic background.
In order to accumulate statistics, the PDF for each of these variables are generated using events in which $^{12}$B is produced by muons, after confirmation of the agreement with those from $^{9}$Li.
Li veto rejects 55\% of the cosmogenic $^{9}$Li and $^{8}$He background.
The IBD inefficiency is measured to be (0.508 $\pm$ 0.012)\,\% by counting IBD candidates in coincidence with off-time muons.

\vspace{1mm}
Muons interacting in the surrounding material can produce multiple fast neutrons which can enter the ID producing one or more recoil protons simulating the positron and some being captured and producing the delayed coincidence. 
The following cuts have been devised to reduce this correlated background.

{\bf OV veto}: 
Muons (including the ones that stop in the detector) traversing the OV can generate an OV trigger.
IBD candidates are rejected if such a trigger in coincident with the prompt signal within 224\,ns exists. 
Using a fixed rate pulser trigger, the IBD inefficiency due to the OV veto is calculated to be 0.056\%.

{\bf IV veto}: 
Extending its original function of rejecting muons, the IV is used in the analysis to tag and reject FN, remaining SM and accidental backgrounds.
IV tagged events are those triggered by the ID energy deposition but exhibiting energy deposition in the IV detector within the same FADC window, i.e. effective $< 256\,$ns time coincidence and threshold-less IV readout.
The implementation rationale of the IV veto definition is similar to that of the n-Gd analysis\cite{ref:DCIII_nGd}, but with major improvements specific to the n-H capture sample.
IBD candidates are IV-tagged and rejected if either or both of the prompt and delayed signals satisfy the following requirements: IV PMT hit multiplicity $\geq$ 2 (where a PMT hit is defined as $\gtrsim 0.2\,$p.e.), energy deposition in the IV $\gtrsim$ 0.2\,MeV, energy depositions in the IV and ID reconstructed within 4.0\,m in space and 90\,ns in time of each other.
Despite the fact that the IV, being the outermost layer, is exposed to a large rate ($>100\,{\rm ks^{-1}}$) of surrounding rock radioactivity, threshold-less PMT signal recording by the IV FADC allows to observe such small, 2 PMT hit, signals caused by energy deposition in the IV by $\gamma$ and fast neutrons from surrounding rock.
The last three conditions are designed to suppress inefficiency of IBD signals due to accidental coincidence by radioactivity.
Following these conditions, the IV veto was found to introduce no IBD inefficiency with a systematic uncertainty of 0.169\%.

\begin{figure}
\begin{center}
	\includegraphics[height=63mm]{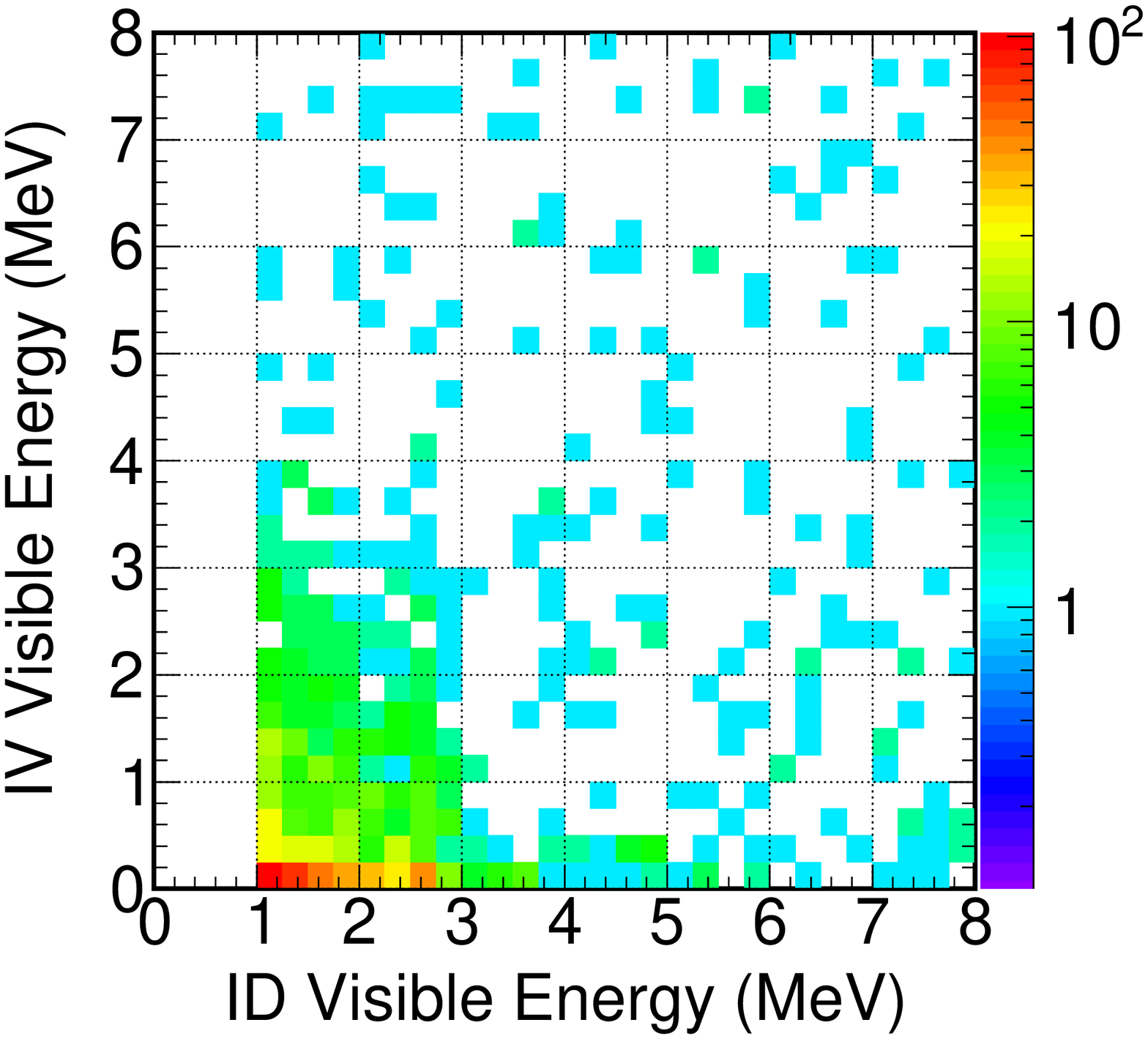}
	\includegraphics[height=63mm]{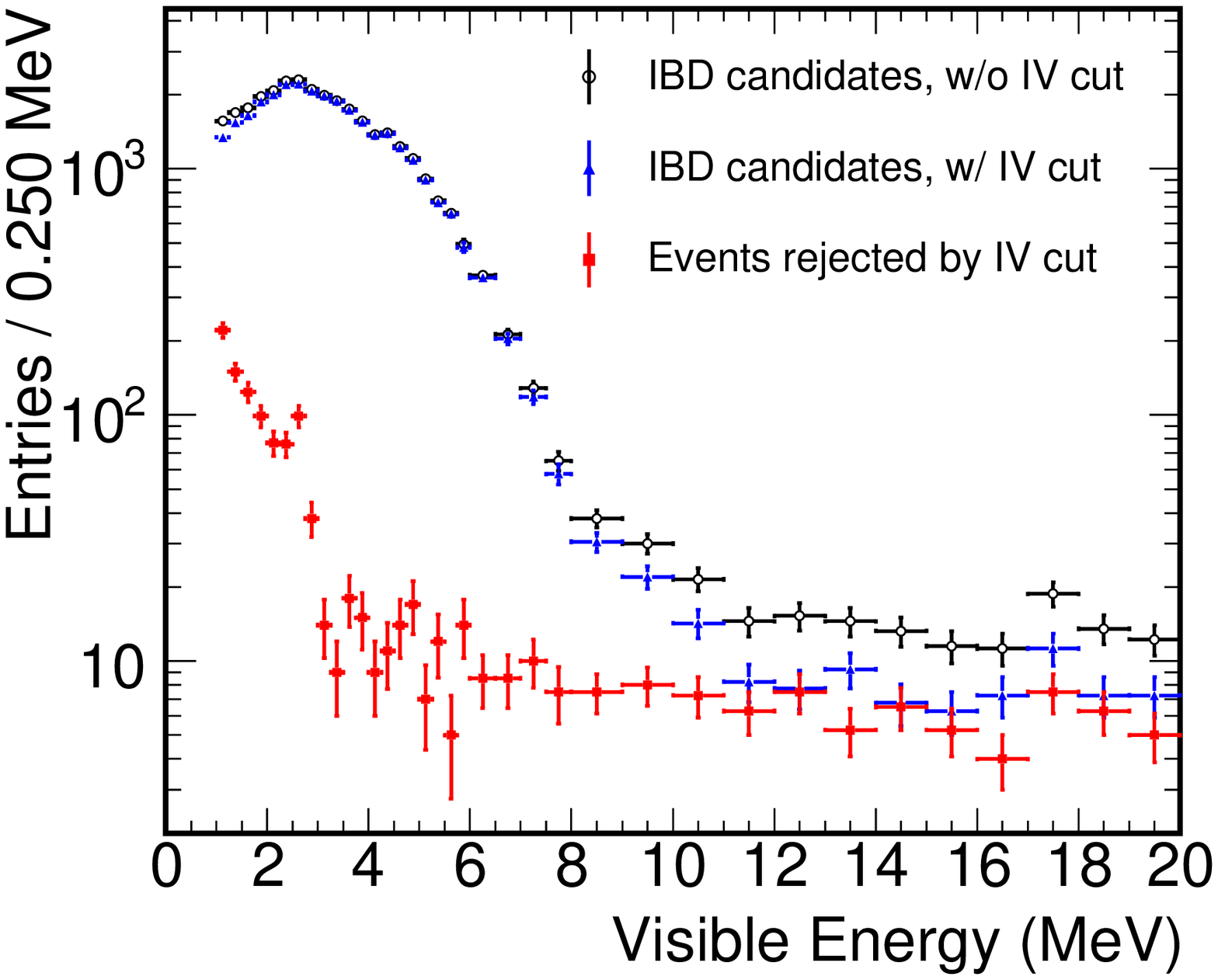}
	\caption{Left: Correlation of the prompt visible energies observed in the ID and IV for events rejected by the IV cut either due to the prompt or delayed event in IBD candidates. The cluster of events with low ID and IV energies up to 3\,MeV in total are interpreted as due to the same $\gamma$'s which deposit energy in both the ID and IV. Uniformly distributed events are due to fast neutrons producing recoil protons in both the ID and IV. Right: Black circle and blue triangle points refer to the prompt energy distributions of IBD candidates before and after the IV cut is applied, respectively. Red square points denote the events rejected by the IV cut.}
	\label{fig:IVV}
\end{center}
\end{figure}

In contrast to the n-Gd analysis, in which the main target was FN background, the IV veto in the n-H analysis rejects a significant amount of the accident backgrounds arising from multiple Compton scattering of $\gamma$'s in the IV and ID.
These $\gamma$  rays are emitted from radioactive nuclei in the surrounding rock and the spectrum shape indicates that 2.6\,MeV $\gamma$'s from $^{208}$Tl are dominant in our delayed energy window.
Figure~\ref{fig:IVV} shows that the majority of IV-tagged events are actually such $\gamma$ Compton events accumulated at low energy.
By applying the IV-tagging to both the prompt and delayed candidates, a total of 27\% of the remaining accidental background after the ANN cut is rejected.

{\bf Multiplicity Pulse Shape (MPS) veto}: 
Recording the waveform of all the PMT signals with a time bin of 2\,ns has allowed the use of a new cut to reduce the FN background based on identifying small energy deposits in the ID, which can be due to other recoil protons before the main signal in the same FADC window.
For this analysis, the start times of all pulses in an event are extracted from the waveform by the same algorithm as in Ref.~\cite{ref:oP} and accumulated, after correcting for different flight paths, to form the overall MPS of the event.
Zero of the PS distribution is defined as the start time of the first pulse after removal of isolated noise pulses.

\begin{figure}
\includegraphics[width=78mm]{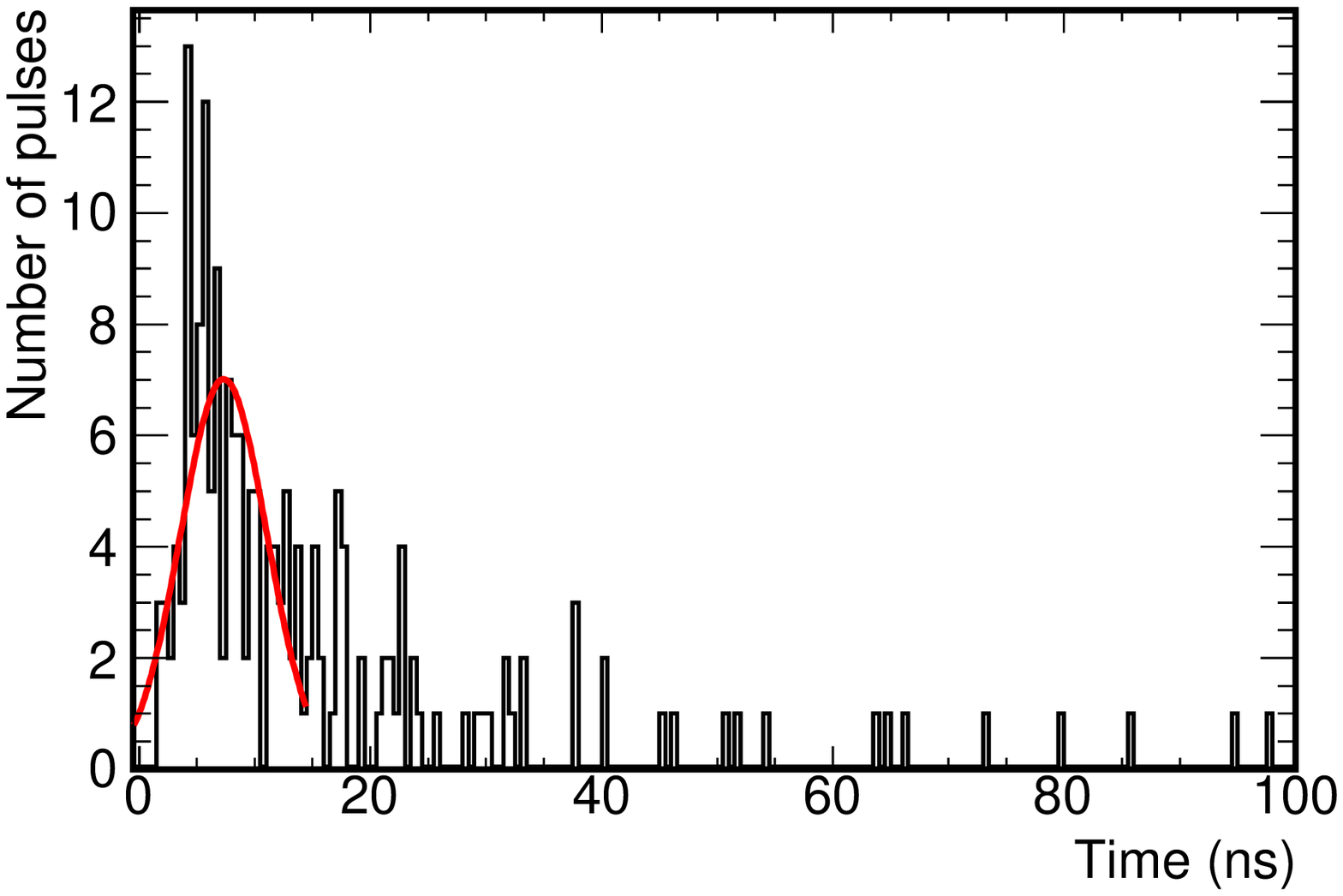}
\includegraphics[width=78mm]{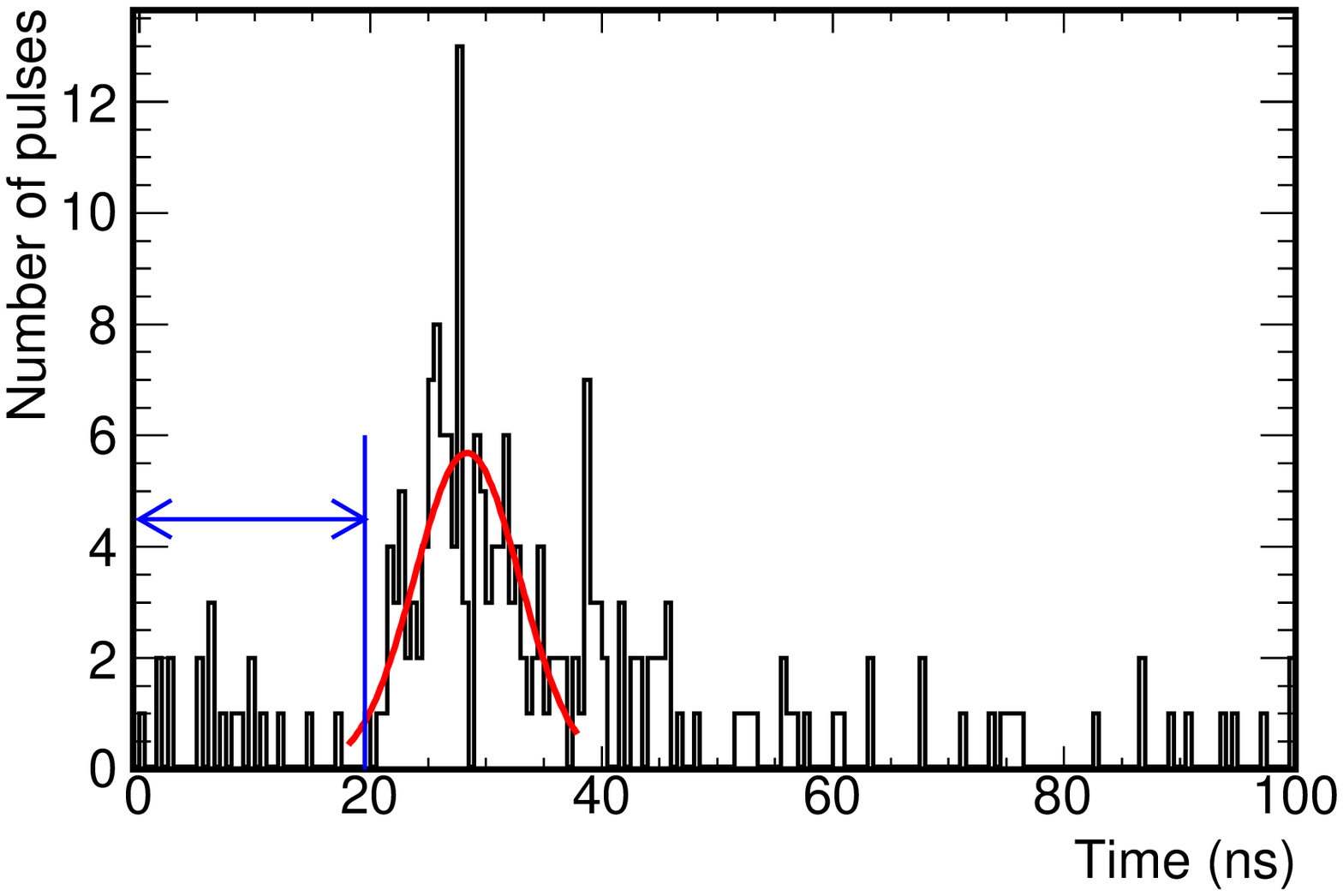}
\caption{The multiplicity pulse shape (MPS) represented as the number of the start times of all pulses in an event, as a function of start time shown for an IBD event (left), and a Fast Neutron (FN) (right). The red curves are the gaussians used to determine the shift described in the text. The blue arrow shows the size of shift which is negative for the IBD event and hence not indicated while a sizable shift due to several pulses before the main signal is visible for the FN candidate. These preceding pulses are understood to be due to multiple recoil protons at different vertices.}
\label{fig:PSexamples}
\end{figure}

MPS are shown in Figure~\ref{fig:PSexamples} for a typical IBD event (left) and a FN event (right).
For the FN, the large cluster of start times is shifted from zero due to other proton recoils from neutrons produced in muon spallation interaction. 
The highest peak in MPS is fit to a Gaussian yielding its mean, $m$, and width, $\sigma$. 
The MPS initial position is defined as $\lambda = m - 1.8 \times \sigma$, as depicted by the blue vertical line in Figure~\ref{fig:PSexamples}. 
The distribution of the shift of $\lambda$ from the start time of the waveform (defined as the time of the first non-isolated pulse) for a $\gamma$ emitter $^{60}$Co source, characteristic of IBD positrons, shows that a cut at 5\,ns on this shift retains all the source events while it rejects a large fraction of FN background.
This cut is not applied to events with prompt energy between 1.2 and 3.0\,MeV recognized as a double-peaked ortho-positronium, oPs, by a dedicated algorithm~\cite{ref:oP} or for events below 1.2\,MeV for which the low energy first peak would not be recognized by the algorithm. 

As the multiple neutron production from spallation interaction by cosmic muon is complicated process and not implemented in the Double Chooz MC, the reduction of the FN contamination by the MPS veto is evaluated using the data with three selections of FN.
The MPS veto rejects $24\pm2$\% of OV tagged events, $29\pm3$\% of IV tagged events  and $27\pm2$\% of IBD selected events with prompt energy larger than 12\,MeV, all consistent within the statistical uncertainties.
Those rejected by the MPS veto display an energy spectrum consistent with the FN background tagged by the IV and OV (see Section~\ref{section:Background}).
The IBD inefficiency of this cut is estimated by studying the events between 1.0 and 20\,MeV with a shift above 5\,ns and occurring in the bottom half of the detector to suppress the FN contribution.
The number of FN in the IBD signal region is calculated by extrapolation from $> 12\,{\rm MeV}$ assuming they are pure FN.
Subtracting this FN estimate from the observed number of events yields a number of IBD events failing the shift cut that is consistent with zero with an uncertainty of 0.1\%.

\section{Residual background estimation}
\label{section:Background}
Methods to reduce the different sources of background have been described in Section~\ref{section:Selection}. 
This section describes how the rate and energy distribution of their remaining contributions are measured by data-driven methods in order to include them in the final fit.

The {\bf accidental background} rate and spectrum shape are measured by searching for delayed events in 200 consecutive time windows starting 1\,s after the prompt candidate, keeping all other criteria unchanged.
The accidental rate is measured to be: $4.334 \pm 0.007(\rm stat) \pm 0.008 (\rm syst)$ events/day after correcting for live-time, muon veto and multiplicity effects affecting differently the on-time and off-time events.
This accidental background rate corresponds to approximately 6\% relative to the predicted IBD signal rate, largely suppressed by the new selection with respect to the previous n-H analysis in which accidental background rate was almost the same as the IBD signal rate.

Contamination from the {\bf cosmogenic isotopes} is evaluated from fits to the time interval between the prompt signal of IBD candidates and the previous muons ($\Delta T_{\mu}$) without the Li veto (see Section~\ref{section:Selection}) and the fraction of vetoed events is subsequently subtracted.
Muons are divided into sub-samples according to their energy in the ID ($E_{\mu}$), as the probability of generating Li increases with $E_{\mu}$.  
After subtraction of the random background determined from a sample of off-time muon-IBD coincidences, the sample above 600\,MeV$^{*}$\footnote{MeV$^{*}$ represents MeV-equivalent scale as the energy reconstruction is not ensured at such high energy due to non-linearity associated with flash-ADC saturation effects.} is the only one that can generate a sufficiently pure sample of Li without applying cuts on the distance ($d$) between the muon and the prompt signal.
The lateral distance profile (LDP) was evaluated by a simple simulation as follows: a) generated muon-IBD coincidences separated by an exponential distribution of $d$ with an averaged distance $\lambda$, b) implemented the reconstruction resolution of the two deposits and c) applied the acceptance of the detector. 
Fitting the resulting LDP to the data yielded a  $\lambda$ of 491\,mm from which acceptance corrected probability density functions (pdf's) of the LDP for each $E_{\mu}$ sub-sample could be generated. 
A Li sample was then obtained from the data, divided into several ranges of $E_{\mu}$ and restricted to coincidences with $0 \leq d \leq d_{\rm max}$. 
The efficiency of the $d_{max}$ cut was evaluated from the generated pdf's.
Several samples were obtained by varying $d_{max}$ between 400 and 1000\,mm, evaluating the Li rate for each sample through a fit of its $\Delta T_{\mu}$ distribution using exponentials describing the cosmogenic decays and a flat background. 
The average and rms of these rates were taken, respectively, as a measure of the Li contribution, $R_{\rm Li}$, and its systematic error: \mbox{ $R_{\rm Li} = 2.76 ^{+0.43}_{-0.39}(stat) \pm 0.23(syst)$} events/day.   

\begin{figure}
\begin{center}
	\includegraphics[width=90mm]{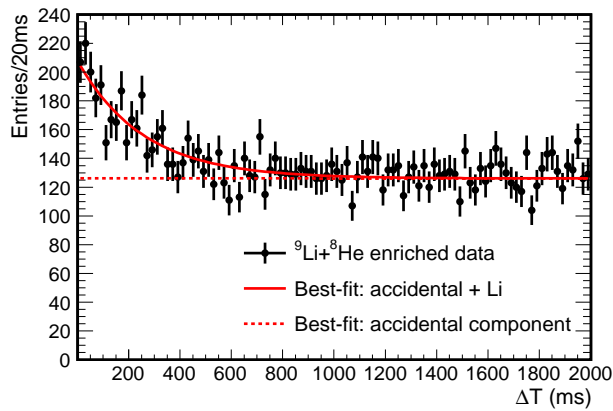}
	\caption{$\Delta T_{\mu}$ distribution of the Li enriched sample, described in the text. The solid red curve shows the best fit and the dashed red curve the accidental component.}
	\label{fig:LiDtMuon}
\end{center}
\end{figure}

As an alternative approach, the minimum contamination of the Li background was estimated by a Li-enriched sample selected as the sum of two samples: 1)$E_{\mu} > 400\,$MeV$^{*}$ and one or more neutron candidates 2) $E_{\mu} > 500\,$MeV$^{*}$, no neutron candidate and $d < 1000\,$mm.
A fit to the resulting  $\Delta T_{\mu}$ distribution, shown in Figure~\ref{fig:LiDtMuon}, gives a minimum Li rate of $2.26 \pm 0.15$\,events/day.
Combining the two measurements described above yields a Li rate of $2.61^{+0.55}_{-0.30}$\,events/day, where the lower bound has been improved by use of the minimum rate. 
The final Li rate is obtained as $2.58^{+0.57}_{-0.32}$\,events/day after including systematics from the LDP, fit configuration and a contribution from $^{8}$He of \mbox{$(7.9 \pm6.6)\,\%$} based on the measurement by KamLAND~\cite{ref:KamLAND_He}, rescaled to our overburden. 

A fit to the $\Delta T_{\mu}$ distribution of events failing the Li veto yielded a Li rate of $1.63 \pm 0.06$\,events/day rejected by this cut, a value confirmed by a simple counting approach, in which the number of Li candidates in the off-time windows is subtracted from the number of Li candidates rejected by the Li veto. 
The remaining Li contamination in the IBD sample is $0.95^{+0.57}_{-0.33}$\,events/day. 
The spectrum shape of the $^{9}$Li and $^{8}$He background, used as input to the final fit, is measured from the Li candidate events selected by the Li veto after subtraction of the accidental muon-IBD coincidences obtained in off-time windows.
It is shown in Figure 15 of Ref.~\cite{ref:DCIII_nGd}. 

\begin{figure}
\begin{center}
	\includegraphics[width=110mm]{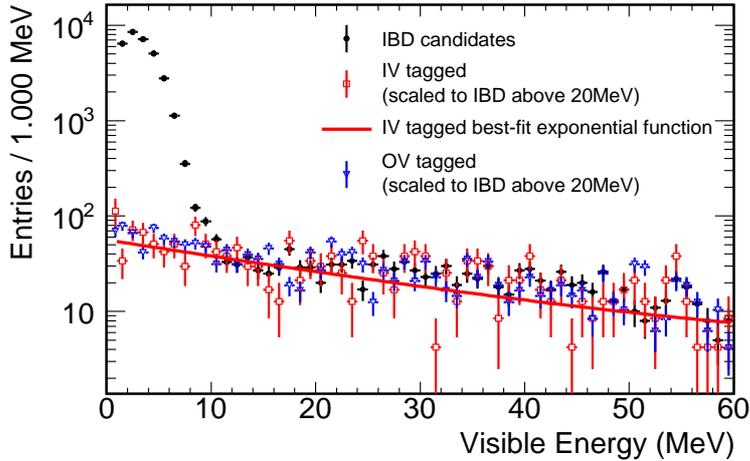}
	\caption{The IBD selection extended to 60 MeV in visible energy (black) together with the Fast Neutron spectrum (red and blue) obtained with an Inner Veto and Outer Veto tag (as explained in the text) normalized to the IBD above 20 MeV. The solid red curve shows the best fit to the Inner Veto tagged events, used to estimate the FN background in the signal region.}
	\label{fig:FNIVT}
\end{center}
\end{figure}

The contribution of {\bf FN and SM background} in the IBD prompt energy range is estimated by measuring the number of FN in that region that are tagged by an FN algorithm and correcting it by the FN tag efficiency.
An IV tag selected events with $E_{\rm IV} > $ 6\,MeV, IV-ID position correlation between 1.1 and 3.5\,m and time correlation within 60\,ns.
The efficiency of this tagging is measured to be (23.6 $\pm$ 1.5)\,\% using events with energy greater than 20\,MeV which are assumed to be a pure FN sample.
Using an extended IBD event sample with prompt energy up to 60\,MeV the tagged FN contamination was measured and fitted using an exponential function yielding $dN/dE_{vis} = p_{0} \times \exp(-p_{1} \times E_{\rm vis}) + p_{2}$, with $p_{0} = 12.52/{\rm MeV}$, $p_{1} = 0.042/{\rm MeV}$ and $p_{2} = 0.79/{\rm MeV}$.
Integrating this curve over the prompt energy window and correcting for the tagging efficiency resulted in an FN contribution of $1.55 \pm 0.15$ events/day.
This function normalized to this rate was used as input to the final fit together with the uncertainties on the fit parameters and their correlation.
A consistent rate and spectrum shape of the FN background was obtained by a muon tagging method, based on the OV, using events that passed all the IBD selection criteria, except the OV veto, and were tagged by the OV.
The estimate based on the IV tagging is used in the neutrino oscillation fit as it tags FN background from all directions and the IV has been in operation for the entire data taking period.
Figure~\ref{fig:FNIVT} shows the visible energy spectrum of IBD candidates extended to 60 MeV and of IV and OV tagged events normalized to the IBD events above 20 MeV.
The fit function to IV tagged events is overlaid.
We observed a rate of FN background selected with n-H captures, mostly in the GC, that decreases with increasing energy, unlike the flat energy spectrum of FN background observed with n-Gd capture in NT.

A contamination of SM in the final IBD sample is estimated using a sample of events passing the IBD cuts except that they are coincident with an OV trigger. 
SM occur mostly in the chimney and they are identified through the difference between two vertex reconstruction log likelihoods: one using the standard reconstruction vertex and a second one, which tends to be smaller for SM, computed using an assumed vertex position in the chimney. 
The contribution of SM is estimated to be 0.02 events/day which is included in the FN and SM background rate and spectrum shape measurements by the IV tag.  

A small contamination of double n-H captures originated from cosmogenic fast neutrons was observed in the IBD candidates.
This contamination arises due to the fact that the preceding recoil protons which would have caused it to be rejected by the multiplicity cut, were not identified.
The rate of less than 0.2\,events/day of this background allowed it to be neglected in the oscillation fit.

Contamination of correlated light noise background, caused by two consecutive triggers due to light noise, was identified in our previous n-H analysis~\cite{ref:DCII_nH}.
This background is fully rejected with the new light noise cuts used in this paper.

These estimated background rates are summarized in Table~\ref{table:BGSummary} together with those from our previous analysis~\cite{ref:DCII_nH} and are used as inputs to the neutrino oscillation fit described in Section~\ref{section:Analysis}.

\begin{table}[ht]
\begin{center}
	\begin{tabular}{|c|c|c|}
	\hline
	Background & H-III (d$^{-1}$)& H-II (d$^{-1}$)\\
	\hline
	Accidental & $4.33 \pm 0.01$ & $73.45 \pm 0.16$\\
	Cosmogenic $^{9}$Li/$^{8}$He &$0.95^{+0.57}_{-0.33}$ & $2.8 \pm 1.2$\\
	Fast-n + Stopping muons & $1.55 \pm 0.15$  & $3.17 \pm 0.54$\\
	\hline
        Total & 6.83$^{+0.59}_{-0.36}$   & 79.4$\pm$1.3    \\
	\hline
	\end{tabular}
	\caption{Summary of background estimates used in this analysis, H-III, and in H-II our previous hydrogen capture publication~\cite{ref:DCII_nH}.}  
	\label{table:BGSummary}
\end{center}
\end{table}

\section{Detection Systematics}
\label{section:Efficiency}
To account for slight differences between the data and the treatment of the MC simulation, a correction factor to the normalisation of the MC prediction is computed.
Three correction factors account for the detection of neutron from IBD signals: $c^{\rm H}$ corrects for the fraction of neutron captures on H; $c^{\rm Eff}$ corrects for the neutron detection efficiency; and $c^{\rm Sio}$ corrects for the modeling of \mbox{spill in/out} by the simulation.
A fourth factor corrects for the number of free protons in the detector which is associated with the IBD interaction rate.
Each factor and its systematic uncertainty is described in this section.

In the NT, neglecting the 0.1\% fraction of captures on carbon, the H fraction is the complementary value of the gadolinium fraction computed for~\cite{ref:DCIII_nGd} yielding a correction factor of $c_\textrm{NT}^\textrm{H}=1.1750\pm0.0277$ including both statistical and systematic uncertainties. 
In the GC, the hydrogen fraction is measured using a $^{252}$Cf neutron source located at the upper edge of the GC cylindrical vessel (far from the NT) to avoid Gd captures. It is defined as the ratio of the number of captured neutrons yielding a visible energy between 0.5 and 3.5\,MeV to those in an energy range extended to 10\,MeV.   
Based on three source deployments and their simulation, the correction factor is found to be: $c_\textrm{GC}^\textrm{H}=1.0020\pm0.0008$ including the systematic uncertainty evaluated by varying the low energy threshold from 0.5 to 1.5\,MeV.
This factor has been checked to be consistent with the value obtained using neutrons from IBD events spread over the whole volume.
Combining $c_\textrm{NT}^\textrm{H}$ and $c_\textrm{GC}^\textrm{H}$, the correction factor over the full volume is obtained as: $c^\textrm{H}=1.0141\pm0.0021$.

The detection efficiency of neutron captures is measured using IBD candidates observed over the whole detection volume, NT and GC, and, to limit the background, using more restrictive cuts on the prompt signal: $1.0 < E_{\rm vis} < 9$\,MeV; and $F_{\rm V} < 5.8$. The remaining accidental background is measured and accounted for using off-time coincidences.
The capture efficiency is then defined as the ratio of the number of IBD candidates selected by the standard delayed signal window to that selected by an extended one: $ANN > -0.40$; $0.25 < \Delta T < 1000\,\mu{\rm sec}$; $\Delta R < 1.5\,{\rm m}$; and $1.3 < E_{\rm delayed} < 3.1\,{\rm MeV}$.
The discrepancy of the efficiency between the data and MC is found to be ($0.05 \pm 0.17$)\,\%, where the uncertainty includes a statistical component (0.13\%), a contribution from the accidental correction factor (0.01\%) and a systematic uncertainty (0.11\%), estimated as the change in the correction when only IBD candidates in the lower half of the detector are used. Since no significant discrepancy is observed, the correction factor is taken as $c^{\rm Eff} = 1.0000 \pm 0.0022$. A consistent number is obtained using Cf source data.

Particles produced in the detector can propagate in or out of a given detector volume. Spill effects are predominantly affected by neutron modeling, itself dependent on the treatment of molecular bonds between hydrogen and other atoms, implemented through a patch in our Geant4 simulation.
To estimate the spill systematic uncertainty we have compared Geant4 to another simulation~\cite{ref:TRIPOLI4}, \mbox{TRIPOLI-4}, known for its accurate modeling of low energy neutron physics.
Since \mbox{TRIPOLI-4} does not include radiative photon generation and scintillation light production and propagation, for each \mbox{TRIPOLI-4} event the visible delayed energy and the prompt to delayed distance were built based on Geant4 distributions.
Events were generated in all detector volumes and the number of prompt events in each volume in \mbox{TRIPOLI-4} was normalized to match that in Geant4. 
After propagating the positron and neutron the number of spill events in the two simulations differed by 0.18\% of the total number of generated events, a measure of the spill uncertainty.
The possible inadequacy of Geant4 distributions to apply to \mbox{TRIPOLI-4} events introduced an additional 0.22\% uncertainty.
Systematic uncertainties associated with the energy scale and statistical uncertainties of the simulations are found to be 0.07\% and 0.03\%, respectively. 
Taken together, these uncertainties gave a total spill uncertainty of 0.29\% and a correction factor of $c^{\rm Sio} = 1.0000 \pm 0.0029$. 

Combining $c^{\rm H}$, $c^{\rm Eff}$, $c^{\rm Sio}$, the final MC correction factor accounting for the neutron detection efficiency is: $1.0141 \pm 0.0042$.

The number of free protons in the detector introduces an additional correction factor of 1.0014$\pm$0.0091, which is currently the dominant systematic uncertainty associated with the IBD signal detection.
The uncertainty arises mostly from the GC, which was originally not considered as a target for IBD interactions, and hence affects the detection of n-H capture signals.
The proton number uncertainty in the GC includes the contributions of the mass estimation from a geometrical survey of the acrylic vessels combined with liquid density measurements and the hydrogen fraction determination in the GC scintillator.
Among these, the uncertainty is dominated by the measurement of the hydrogen fraction, which was determined using elemental analysis of the liquid mixture.
The analysis of the organic material is based on the method of combustion and consists of three phases: purge, burn and analyze. 
First, the sample and all lines are purged of any atmospheric gases. 
During the burn phase, the sample is inserted into the hot furnace and flushed with pure oxygen for very rapid combustion. 
In the analyze phase, the combustion gases are measured for carbon, hydrogen and nitrogen content with dedicated detectors.
This uncertainty is dominant in the current n-H analysis using only the FD, but can be reduced in the comparison of ND and FD in near future.

Total MC normalisation correction factors including other sources are summarized in Table~\ref{table:MCCF} with the uncertainties.

\begin{table}[ht]   
  \begin{center}
    \begin{tabular}{| c|c|c |}
      \hline
      Correction source & MC correction & Uncertainty(\%) \\
      \hline
       DAQ \& Trigger & 1.000 & $< 0.1$ \\
       Veto for 1.25\,ms after muons & 0.940 & $< 0.1$ \\
       IBD selection & 0.979 & 0.2 \\
       F$_{\rm V}$, IV, OV, MPS, Li vetoes & 0.993 & 0.2 \\
       H fraction & 1.014 & 0.2 \\
       Spill in/out & 1.000 & 0.3 \\
       Scintillator proton number & 1.001 & 0.9 \\
       \hline
       Total  & 0.928 & 1.0 \\
       \hline
   \end{tabular}
  \end{center}
  \caption{Summary of inputs for the MC normalisation correction factor and their uncertainties. IBD selection includes the correction for IBD inefficiency due to multiplicity condition (Section~\ref{section:Selection}). Inefficiencies due to each background veto are summarized in Table~\ref{table:MCIneffCF}.}
  \label{table:MCCF}
\end{table}

\section{Neutrino Oscillation Analysis}
\label{section:Analysis}
Applying the selection cuts described in section~\ref{section:Selection} yielded 31835  IBD candidates in 455.57 live days with at least one reactor operating. 
Given the overall MC correction factor of 0.928 $\pm$ 0.010 (see Table~\ref{table:MCCF}), the corresponding prediction of expected events from the non-oscillated neutrino flux is 30090$\pm$610 and a background of \mbox{$3110^{+270}_{-170}$} as listed in Table~\ref{table:IBDcandidates}.
In addition Double Chooz observed 63 events in 7.15 days of data during which both reactors were off and in which the number of residual reactor $\overline\nu_e$ is evaluated by a dedicated simulation study~\cite{ref:DC_Off} to be \mbox{2.73 $\pm$ 0.82 events.} 
Including the estimated background, the total number of expected events in this reactor off running is \mbox{$50.8^{+4.4}_{-2.9}$}, consistent with the number of events observed, thus validating our background models. 
This measurement is used to constrain the total background rate in the neutrino oscillation analyses. 
Uncertainties on the signal and background normalisation are summarized in Table~\ref{table:RateError}.

\begin{figure}
\includegraphics[height=62mm]{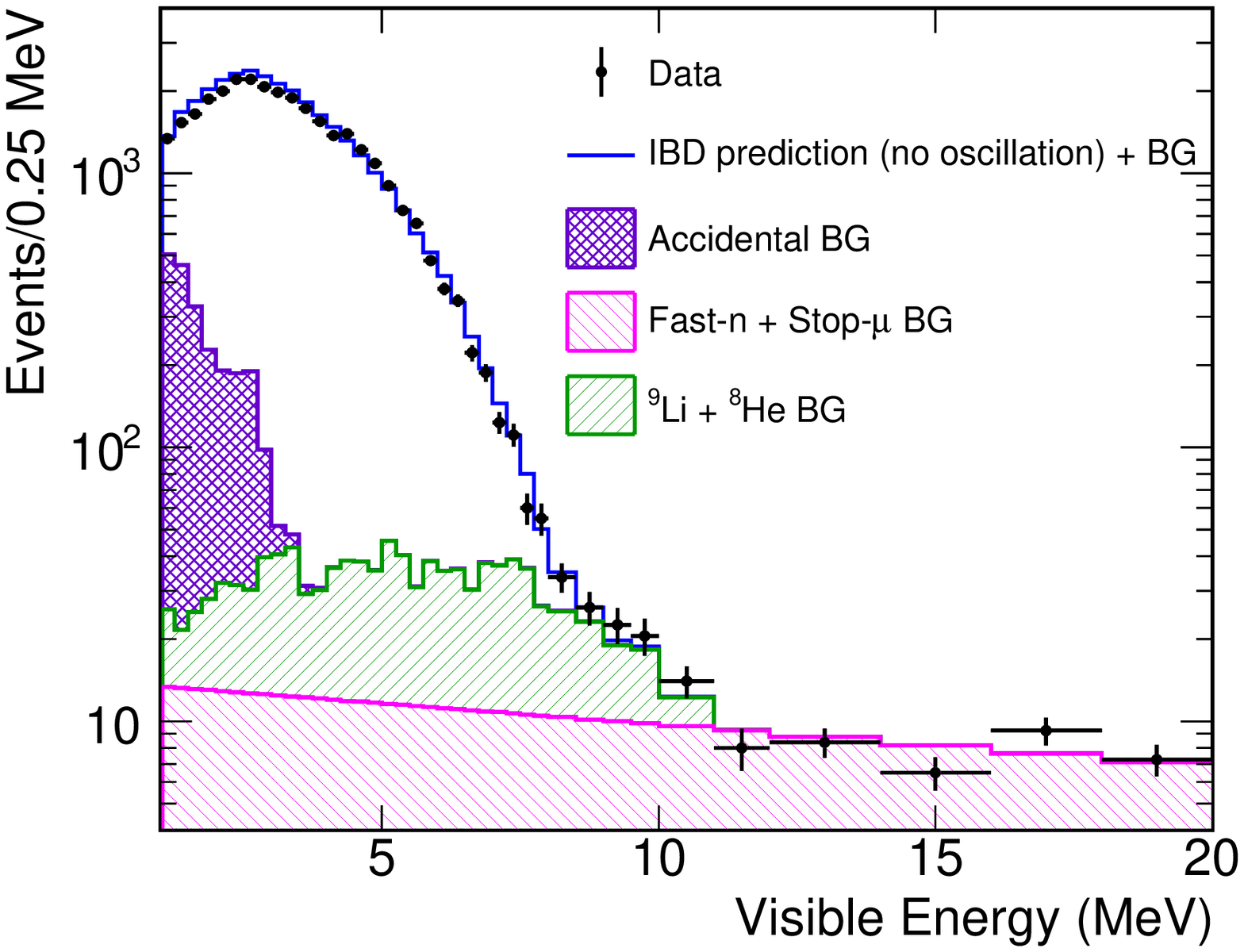}
\includegraphics[height=62mm]{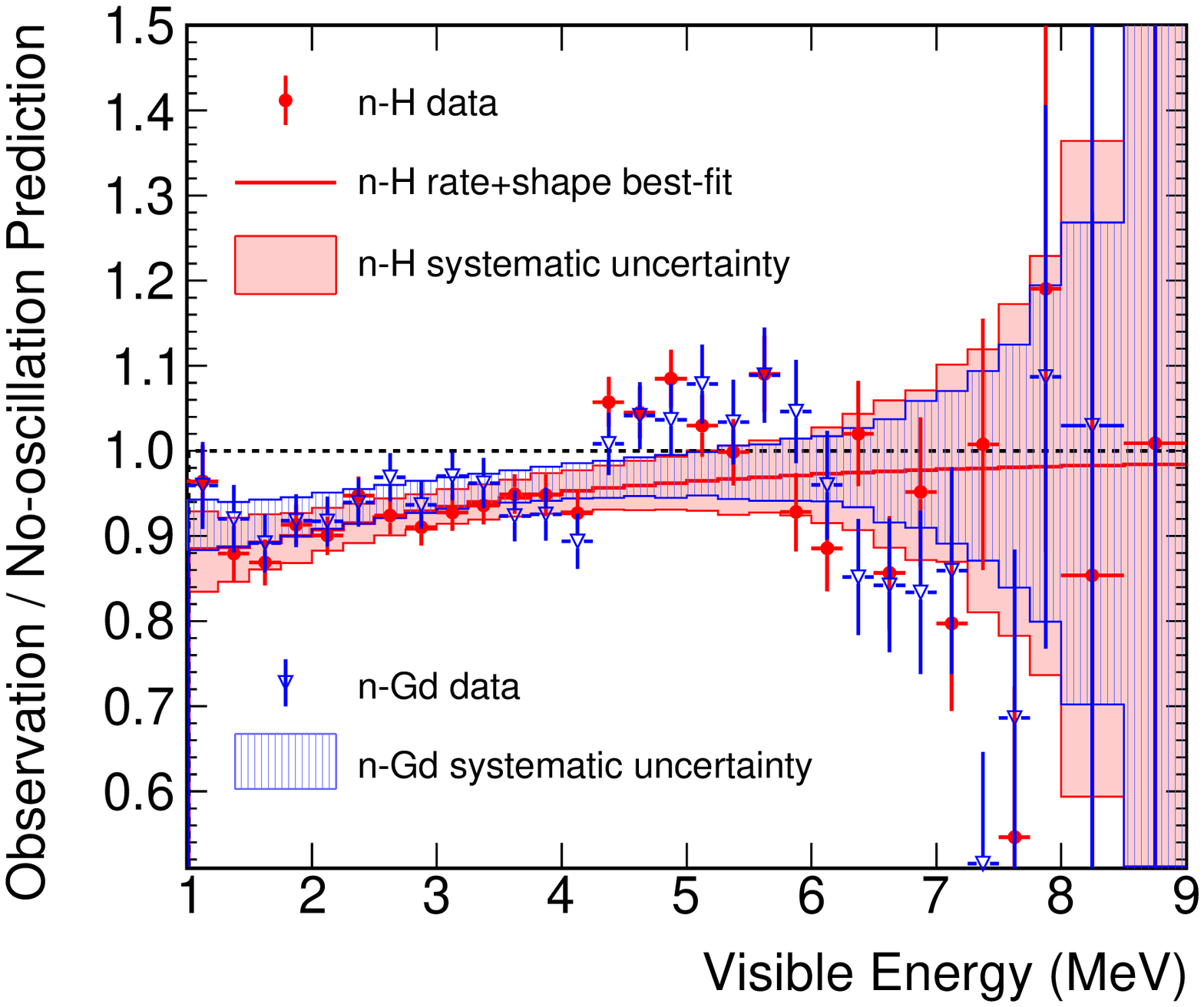}
\caption{\label{fig:Spectrum}Left: The visible energy spectrum of IBD candidates (black points) compared to a stacked histogram (blue) of the expected IBD spectrum in the no-oscillation hypothesis, the accidental (purple), $^9$Li + $^8$He (green) and the fast neutron (magenta) background estimates.
Right: The ratio of the IBD candidates visible energy distribution, after background subtraction, to the corresponding distribution expected in the no-oscillation hypothesis. 
The red points and band are for the hydrogen capture data and its systematic uncertainty described in this publication and the blue points and band are from the Gd capture data described in Ref.~\cite{ref:DCIII_nGd}. 
Red solid line show the best fit from the R+S analysis.}
\end{figure}

Figure~\ref{fig:Spectrum} (left) shows the visible energy spectrum of the IBD candidates together with the expected IBD spectrum in the no-oscillation hypothesis augmented by the estimates of the accidental and correlated background components. 
The background components are also shown separately in the figure.  
A deficit of events is obvious in the region affected by $\theta_{13}$ oscillations. 
Figure~\ref{fig:Spectrum} (right) shows the ratio of the data, after subtraction of the backgrounds described in Section~\ref{section:Background}, to the null oscillation IBD prediction as a function of the visible energy of the prompt signal.
In addition to the energy dependent deficit seen in the data below 4\,MeV, the same spectrum distortion is observed above 4\,MeV characterized by an excess around 5\,MeV, as was observed in the equivalent ratio obtained in neutron captures on Gd~\cite{ref:DCIII_nGd}, also shown in the figure.

Interpreting the observed deficit of IBD candidates as $\overline{\nu}_{e}$ disappearance due to neutrino oscillation allows the extraction of $\theta_{13}$ in a two-neutrino flavour scenario as described by Eq.~\ref{eq:oscillation}. 
Two complementary analyses, referred to as {\bf Reactor Rate Modulation (RRM)} and {\bf Rate+Shape (R+S)} are performed.
The RRM analysis is based on a fit to the observed IBD candidate rate as a function of the predicted rate, which, at any one time, depends on the number of operating reactor cores and their respective thermal power with an offset determined by the total background rate~\cite{ref:DCII_RRM}. 
As explained in section~\ref{section:Experiment}, the normalisation of the reactor flux is constrained by the Bugey4 measurement~\cite{ref:Bugey4}. 
The precision of the RRM analysis is improved by including the reactor-off data.
The R+S analysis is based on a fit to the observed energy spectrum in which both the rate of IBD candidates and their spectral shape are used to constrain $\theta_{13}$ as well as the background contributions, the latter by extending the fitted spectrum well above the IBD region.
Impact of spectrum distortion to $\theta_{13}$ is found to be negligible within the current precision as described in Section~\ref{section:RS}, although the source of the distortion is not yet understood.

Among the two analyses, as the RRM fit is robust against the spectrum distortion with a constraint from Bugey4, a combined analysis with the gadolinium capture data was carried out based on the RRM fit as in Ref.~\cite{ref:DCII_RRM} and quoted as the primary results, while the spectrum distortion will be further studied at short distance with the near detector now in operation.

\begin{table}
\begin{center}
	\begin{tabular}{|c|c c|c c|}
	\hline
	& \multicolumn{2}{c|}{Reactor On} & \multicolumn{2}{c|}{Reactor Off}\\
	\hline
	Live-time (days) & \multicolumn{2}{c|}{455.57} & \multicolumn{2}{c|}{7.15}\\
	\hline
	IBD Candidates & 31835 & (69.9/day) & 63 & (8.8/day) \\
	\hline
	\hline
	Reactor  $\overline\nu_e$ prediction & $30090 \pm 610$ & $(66.0 \pm 1.3)$ & $2.73 \pm 0.82$ & $(0.38 \pm 0.11)$\\
	Accidental BG & $1974.4 \pm 4.8$ & $(4.33 \pm 0.01) $ & $30.88 \pm 0.40$ & $(4.32 \pm 0.06)$\\
	Cosmogenic $^{9}$Li/$^{8}$He BG & $430^{+260}_{-150}$ & $(0.95^{+0.57}_{-0.33})$ & $6.8^{+4.1}_{-2.4}$ & $(0.95^{+0.57}_{-0.33})$\\
	Fast-$n$ and Stop-$\mu$ BG & $706 \pm 68$ & $(1.55 \pm 0.15)$ & $10.4 \pm 1.4$ & $(1.45 \pm 0.20)$\\
	\hline
	Total estimation & $33200^{+660}_{-630}$ & $(72.9 \pm 1.4)$ & $50.8^{+4.4}_{-2.9}$ & $(7.10^{+0.62}_{-0.41})$ \\
	\hline
	\end{tabular}
	\caption{Summary of observed IBD candidates with the prediction of reactor neutrino signal and estimation of background. Numbers in parentheses show the event rate per day. Neutrino oscillation is not included in the prediction. Background rates in reactor off period were separately measured by the corresponding data except for cosmogenic Li and He background.}
	\label{table:IBDcandidates}
\end{center}
\end{table}

\begin{table}
\begin{center}
	\begin{tabular}{|c|c|c|}
	\hline
	Source & H-III Uncer. (\%) & H-II Uncer. (\%) \\
	\hline
	Reactor Flux & 1.7 & 1.8\\ 
	Statistics & 0.6 & 1.1\\
	Detection Efficiency & 1.0 & 1.6 \\
        Energy scale & 0.1 & 0.3 \\
	$^{9}$Li + $^{8}$He BG & $+ 0.86 / -0.50$ & 1.6\\
	Fast-$n$ and Stop-$\mu$ BG & 0.2 & 0.6\\
	Accidental BG & $< 0.1$ & 0.2\\
	\hline
	Total & $+2.3/-2.2$ & 3.1 \\
	\hline
	\end{tabular}
	\caption{Summary of signal and background normalisation uncertainties relative to the signal prediction. H-III and H-II refer the hydrogen capture analysis in this paper and our earlier publication~\cite{ref:DCII_nH}. Small difference of the flux uncertainty is due to different fuel compositions in the data taking periods. Statistical uncertainty includes the propagation of uncertainty due to accidental background subtraction which is suppressed in H-III analysis with much smaller background contamination than H-II analysis. Energy scale in H-III represents the uncertainty associated with the prompt energy window while the uncertainty on the neutron detection is included in the detection efficiency.}
	\label{table:RateError}
\end{center}
\end{table}

\subsection{Rate + Shape analysis}
\label{section:RS}
This analysis compares the energy spectrum of the observed IBD candidates to the summed spectrum of the estimated background and the expected $\overline{\nu}_{e}$ rate including the oscillatory term introduced in the simulation of the two reactor fluxes as a function of $E_{\nu}/L$. The spectra are divided into 38 bins in visible energy spaced between 1.0 and 20\,MeV.
Extending the spectra to 20\,MeV, well beyond the range of IBD events, allows the statistical separation of the reactor $\overline\nu_e$ signals from the background through their different spectral shapes, thus improving the precision of the background contribution.
The background spectral shapes are measured by the data as described in Section~\ref{section:Background} and the uncertainties in the shapes and in the rate estimates are taken into account in the fit.
The definition of the $\chi^2$ used in the fit to extract \mbox{$\sin^{2}2\theta_{13}$} is described in detail in Ref.~\cite{ref:DCIII_nGd}. 
The value of $\Delta m^{2}$ is taken as \mbox{$2.44^{+0.09}_{-0.10} \times 10^{-3}\, {\rm eV}^2$} from the measurement of the MINOS experiment and assuming normal hierarchy~\cite{ref:MINOS_dm2}.
Correction for the systematic uncertainty on the energy scale is given by a second-order polynomial as: \mbox{$\delta(E_{\rm vis}) = \epsilon_{a} + \epsilon_{b} \cdot E_{\rm vis} + \epsilon_{c} \cdot E_{\rm vis}^{2}$}, where $\delta(E_{\rm vis})$ refers to the variation of the visible energy.
Uncertainties on $\epsilon_{a}$, $\epsilon_{b}$ and $\epsilon_{c}$ are given as $\sigma_{a} = 0.067\,{\rm MeV}$, $\sigma_{b} = 0.022$ and $\sigma_{c} = 0.0006\,{\rm MeV}^{-1}$.
A separate term in the $\chi^{2}$ accounts for the reactor-off contribution, but, because of its low statistics, only the total number of IBD candidates is compared with the prediction.

The best fit with  $\chi^{2}_{\rm min}/d.o.f. = 69.4/38$, is found at \mbox{$\sin^{2}2\theta_{13} = 0.124^{+0.030}_{-0.039}$}, where the error is given as the range which gives $\chi^{2} < \chi^{2}_{\rm min} + 1.0$.
This value is consistent with the RRM measurements of $\sin^{2}2\theta_{13}$ reported in the following sections. 
As expected, the large value of $\chi^{2}$ is due primarily to the 4.25-5.75\,MeV region. 
Excluding the points in this region, as well as their contributions through correlations with other energy bins via the covariance matrix, reduces the $\chi^{2}$ to 30.7 for 32 d.o.f..
In order to examine the impact of the spectral distortion to the measured $\theta_{13}$ value, a test R+S fit was carried out with narrower prompt energy window between 1.0 and 4.0\,MeV.
The variation of $\sin^{2}2\theta_{13}$ was well within 1-$\sigma$ of the measured uncertainty.
The input and output best-fit values of the fit parameters and their uncertainties are summarized in Table~\ref{table:BestFit}, demonstrating the reduction in the uncertainties achieved by the fit.
The ratio of the best fit oscillation prediction to the no-oscillation prediction is shown in the right-hand plot in Figure~\ref{fig:Spectrum}.

\begin{table}[ht]
  \begin{center}
    \begin{tabular}{| l | c | c |}
      \hline
      Fit Parameter & Input Value & Best-Fit Value \\
      \hline
      Accidental BG (d$^{-1}$) & $4.33\pm0.011$ & $4.33\pm0.011$ \\
      Li+He BG (d$^{-1}$) & $0.95^{+0.57}_{-0.33}$ & $1.60^{+0.21}_{-0.24}$ \\
      Fast-$n$ + Stop-$\mu$ BG (d$^{-1}$) & $1.55\pm0.15$ & $1.62\pm0.10$ \\
      Residual $\overline{\nu}_e$ & $2.73\pm0.82$ & $2.81\pm0.82$ \\
      $\Delta m^2\ (10^{-3}$ eV$^2$) & $2.44^{+0.09}_{-0.10}$ & $2.44^{+0.09}_{-0.10}$ \\
      E-scale $\epsilon_{a}$ (MeV) & $0\pm0.067$ & $-0.008^{+0.028}_{-0.020}$  \\
      E-scale $\epsilon_{b}$ & $0\pm0.022$ & $-0.007^{+0.007}_{-0.009}$ \\
      E-scale $\epsilon_{c}$ (MeV$^{-1}$) & $0\pm0.0006$ & $-0.0005^{+0.0006}_{-0.0005}$ \\
      FN shape $p_0$ (MeV$^{-1}$)  & $12.52\pm1.36$ & $12.33\pm1.34$  \\
      FN shape $p_1$ (MeV$^{-1}$)& $0.042\pm0.015$ & $0.037^{+0.015}_{-0.013}$ \\
      FN shape $p_2$ (MeV$^{-1}$) & $0.79\pm1.39$ & $0.39^{+1.48}_{-1.30}$ \\
      \hline
   \end{tabular}
  \end{center}
  \caption{Input values of fit parameters with their estimated uncertainties, compared to the Rate+Shape fit output best-fit values and their errors.}
  \label{table:BestFit}
\end{table}

\subsection{Reactor Rate Modulation Analysis}
\label{section:RRM}
In the Reactor Rate Modulation (RRM) analysis the neutrino mixing angle $\theta_{13}$ and the total background rate ($B$) can be determined simultaneously from a comparison of the observed ($R^{\rm obs}$) to the expected ($R^{\rm exp}$) rates of IBD candidates as was done in our previous publications~\cite{ref:DCII_RRM, ref:DCIII_nGd}. 
During our data-taking there were three well defined reactor configurations: 1) two reactors were on (referred to as 2-On); 2) one of the reactors was off (1-Off); and 3) both reactors were off (2-Off). 
The data set is divided further into seven bins according to reactor power ($P_{\rm th}$) conditions: one bin in 2-Off period, three bins with mostly 1-Off, and three bins with 2-On. 

Three sources of systematic uncertainties on the IBD rate are considered: IBD signal detection efficiency ($\sigma_{\rm{d}}$=1.0\%), residual reactor-off $\overline\nu_e$ prediction ($\sigma_{{\nu}}$=30\%), and prediction of the reactor flux in reactor-on data ($\sigma_{\rm{r}}$) ranging  from 1.72\% at full reactor power to 1.78\% when one or two reactors are not at full power.
The $\chi^2$ is defined as follows:

\begin{eqnarray}
\label{eq:chi2RRM}
\chi^{2} &=& \displaystyle \sum_{i=1}^{\rm 6} \frac{\left(R_{i}^{\rm obs} - R_{i}^{\rm exp}-B\right)^2}{ (\sigma_{i}^{\rm stat})^2} \nonumber \\
    \displaystyle &+& 2 \left[ N_{\rm off}^{\rm obs} \ln \left( \frac{N_{\rm off}^{\rm obs}}{N_{\rm off}^{\rm exp}} \right) + N_{\rm off}^{\rm exp} - N_{\rm off}^{\rm obs}\right] \nonumber \\
     \displaystyle   &+& \frac{\epsilon_{\rm d}^{2}}{\sigma_{\rm d}^{2}} + \frac{\epsilon_{\rm r}^{2}}{\sigma_{\rm r}^{2}} +  \frac{\epsilon_{\rm \nu}^{2}}{\sigma_{\rm \nu}^{2}}  + \frac{\left(B-B^{\rm exp}\right)^2}{\sigma_{\rm bg}^2}\\
\label{eq:Noff}
     N_{\rm off}^{\rm exp} &=& (R_{\rm off}^{\rm \nu} + B) \cdot T_{\rm off}.
\end{eqnarray}

It consists of three parts.
The first part contains the $\chi^2$ contributions from the six reactor-on combinations with the expected rates varied according to the values of the systematic uncertainties parameters and the sin$^{2}$2$\theta_{13}$ in the fit. 
The second part describes the $\chi^2$ contribution of the 2-off data, in which the expected number of events ($N_{\rm off}^{\rm exp}$) is given by the sum of the residual $\overline{\nu}_{e}$ rate ($R_{\rm off}^{\rm \nu}$) and the background rate multiplied by the live-time ($T_{\rm off}$).
$N_{\rm off}^{\rm obs}$ represents the observed number of IBD candidates in 2-Off period.
 The last part, consists of four terms which apply the constraints to the detection efficiency, reactor flux, residual neutrinos and background systematics fit parameters from their estimates and errors.  
The systematic uncertainty on the reactor flux prediction is considered to be correlated between the bins as its dominant source is the production cross-section measured by Bugey4~\cite{ref:Bugey4}.
The prediction of the total background rate and its uncertainty are given as: \mbox{$B^{\rm exp} =  6.83^{+0.59}_{-0.36}$}\,events/day (see Section~\ref{section:Background}).
  
\begin{figure}
\includegraphics[height=70mm]{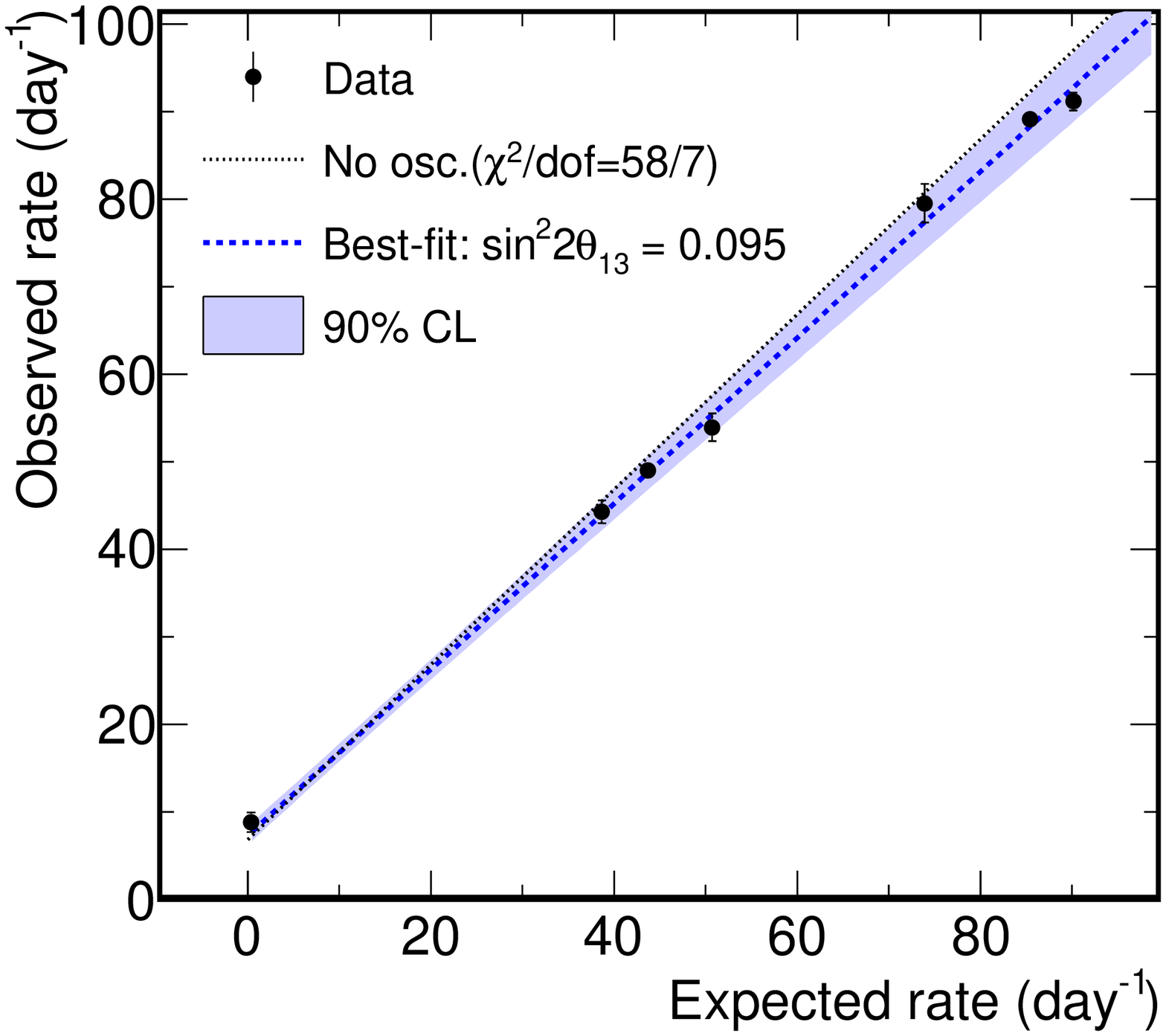}
\includegraphics[height=75mm]{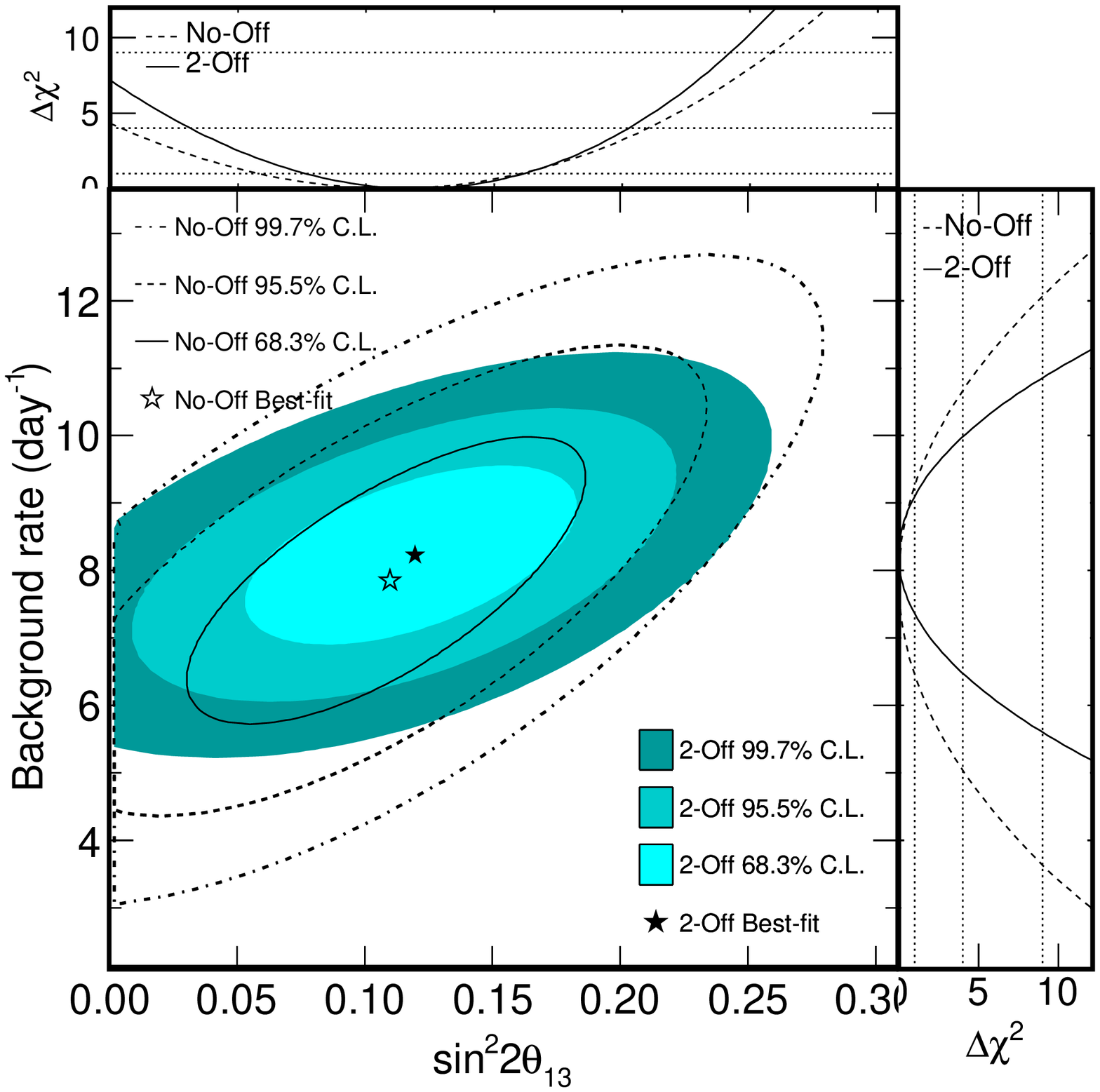}
\caption{\label{fig:RRM} RRM fit results. Left: Observed rate vs reactor flux dependent expected rate and best fit (dotted line) using as input the background estimate and the 2-off data. The dotted line is the no-oscillation expectation. Right: The (background vs $\sin^{2}2\theta_{13}$) 68.3\%, 95.5\% and 99.7\% contours resulting from the RRM fits including (blue) and not including (lines) the 2-Off data sample but not using the background estimate as input. }
\end{figure}

A scan of $\sin^{2}2\theta_{13}$ is carried out minimizing the $\chi^{2}$ with respect to the total background rate and three systematic uncertainty parameters for each value of $\sin^{2}2\theta_{13}$.
The best-fit is for \mbox{$\sin^{2}2\theta_{13} = 0.095^{+0.038}_{-0.039}$} and a total background rate of $B=7.27\pm0.49$\,events/day where the uncertainty is given as the range of $\chi^{2} < \chi^{2}_{\rm min} + 1.0$ with $\chi^{2}_{\rm min}/d.o.f.=7.4/6$. The observed rate is plotted as a function of the expected rate in Figure~\ref{fig:RRM} (left) together with the best fit and no-oscillation expectation.

A  background model independent RRM fit was also carried out by removing the constraint on the total background rate, treating $B$ as a free parameter.
A global scan is carried out on a ($\sin^{2}2\theta_{13}$, $B$) grid minimizing $\chi^{2}$ at each point with respect to the three systematic uncertainty parameters.
The minimum $\chi^{2}$, $\chi^{2}_{\rm min}/d.o.f.= 5.6/5$, is found for $\sin^{2}2\theta_{13} = 0.120^{+0.042}_{-0.043}$ and $B=8.23^{+0.88}_{-0.87}$ events/day, consistent with the RRM fit with background constraint.

Next, the 2-Off term was also removed to test its impact on the precision of the $\theta_{13}$ measurement. The background vs  $\sin^{2}2\theta_{13}$ correlation ellipses are shown in Figure~\ref{fig:RRM} (right). While the central values of the two parameters are hardly changed the uncertainty on $\sin^{2}2\theta_{13}$ is reduced by about 20\% when including the 2-Off data, demonstrating its importance.    

\subsection{Gadolinium and Hydrogen captures combined RRM analysis}
\label{section:Combined}
The RRM fit was then applied to the combined hydrogen capture data presented here and the gadolinium capture data of Ref.~\cite{ref:DCIII_nGd}, including background estimates as input to the fit.
The correlation between the uncertainties of the two data sets were taken as follows: fully correlated for the reactor flux and residual neutrino rate uncertainties and fully uncorrelated for the background uncertainties and the detection systematics.
The result was  $\sin^{2}2\theta_{13} = 0.088\pm 0.033$ (stat+syst) with a  minimum $\chi^{2}_{\rm min}/d.o.f.= 11.0/13$.
The correlation of the detection systematics between the two data sets exists in the NT, amounting to 30\% of the total (NT+GC) detector mass, which would result in a maximum of 30\% of the uncertainty to be fully correlated.
This number is conservative as the dominant component of the detection systematics in the hydrogen analysis is the number of protons in the GC (see Table~\ref{table:MCCF}).
Assuming this hypothesis resulted in a negligible variation in the value of $\sin^{2}2\theta_{13}$, as did the assumption of full correlation of the background systematics.

\begin{figure}
\includegraphics[width=70mm]{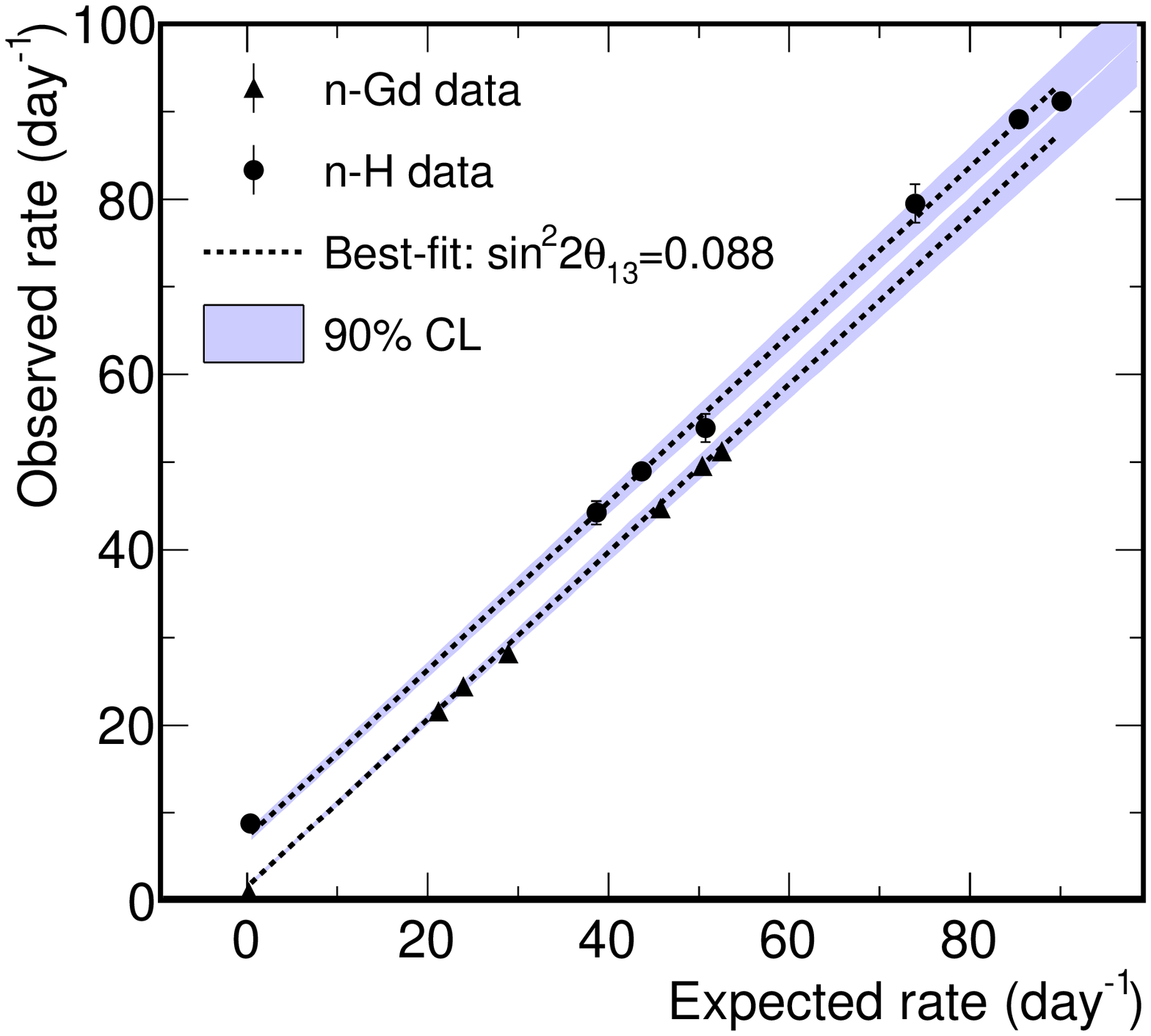}
\includegraphics[width=75mm]{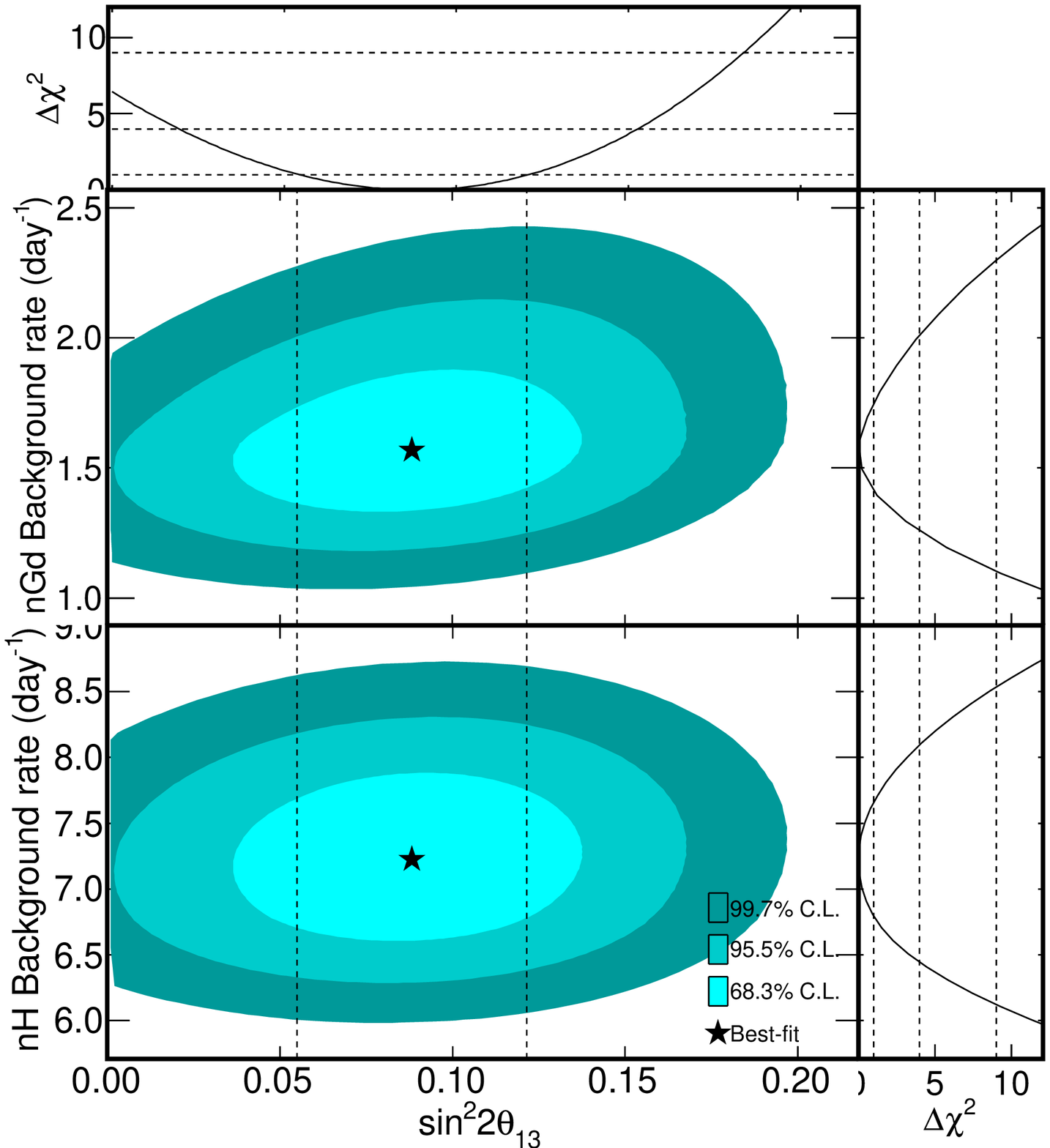}
\caption{\label{fig:RRMComb} Combined RRM fit to the Hydrogen and Gadolinium data sets, assuming no correlations between the background uncertainties of the two data sets and full correlation of the reactor flux and residual neutrinos uncertainties. Left: The observed rate vs the rate expected as a function of reactor power. The fit (dotted lines) is compared to the n-Gd (triangles) and n-H (circles) data sets. Right: The (background vs $\sin^{2}2\theta_{13}$) 68.3\%, 95.5\% and 99.7\% contours resulting from the fit.}
\end{figure}

Figure~\ref{fig:RRMComb} shows the correlation of the observed and expected IBD candidate rates for both data samples together with the combined best-fit and the 68.3\%, 95.5\% and 99.7\% contours on background vs. $\sin^{2}2\theta_{13}$ plane.

\section{Conclusion}
\label{section:Conclusions}
A sample of reactor $\overline{\nu}_{e}$ interactions identified via IBD reactions observed through neutron captures on hydrogen has been used by Double Chooz to measure $\theta_{13}$.
This sample has approximately a factor of 2 more statistics than our previous hydrogen capture publication~\cite{ref:DCII_nH}. 
It is independent of the corresponding sample obtained via neutron captures on gadolinium. 
Several novel background reduction techniques were developed including accidental background rejection based on a neural-network and on a tagging of $\gamma$ Compton scattering in the Inner Veto, and a new cut against fast neutron background using the waveform recorded by the Flash-ADC readout.
These results in a predicted signal to total background ratio of 9.7, a big improvement over the ratio of 0.93 achieved in our earlier hydrogen capture publication.
The systematic uncertainty on the IBD rate measurement was improved from 3.1\% to 2.3\%, of which 1.7\% is associated with the reactor flux prediction.
This was achieved by the reductions of uncertainty on the background estimates, mainly cosmogenic $^{9}$Li $+$ $^{8}$He (from 1.6\% to 0.7\%) and fast neutron $+$ stopping muon (from 0.6\% to 0.2\%), detection systematics (from 1.6\% to 1.0\%) and reduction of statistical uncertainty including accidental background subtraction (from 1.1\% to 0.6\%).

A deficit of events below a visible positron energy of 4\,MeV is consistent with $\theta_{13}$ oscillations whereas a structure above 4\,MeV, described in our earlier publication~\cite{ref:DCIII_nGd}, is an indication for the need for further investigations of the present reactor flux modeling and other systematics effects. 
To be independent of this structure, this publication has focussed on a measurement of $\sin^{2}2\theta_{13}$ based on the event rate as a function of reactor flux (RRM), which does not depend on the shape of the positron energy distribution. 
The analysis, which includes a data sample obtained with both reactors off and uses the background estimates as input, yields a value of \mbox{$\sin^{2}2\theta_{13} = 0.095^{+0.038}_{-0.039}$ (stat+syst)}. 
A cross check of this measurement based on an analysis of the rate + shape of our data results in a consistent value of $\sin^{2}2\theta_{13}$. 
Finally, the RRM method was applied jointly to our hydrogen and gadolinium capture samples resulting in $\sin^{2}2\theta_{13} = 0.088\pm0.033$(stat+syst).

\section{Acknowledgements} 
We thank the French electricity company EDF; the European fund FEDER;
the R\'egion de Champagne Ardenne; the D\'epartement des Ardennes;
and the Communaut\'e de Communes Ardenne Rives de Meuse.
We acknowledge the support of the CEA, CNRS/IN2P3, the computer centre CCIN2P3, and LabEx UnivEarthS in France (ANR-11-IDEX-0005-02);
the Ministry of Education, Culture, Sports, Science and Technology of Japan (MEXT) and the Japan Society for the Promotion of Science (JSPS);
the Department of Energy and the National Science Foundation of the United States;
U.S. Department of Energy Award DE-NA0000979 through the Nuclear Science and Security Consortium;
the Ministerio de Econom\'ia y Competitividad (MINECO) of Spain;
the Max Planck Gesellschaft, and the Deutsche Forschungsgemeinschaft DFG, the Transregional Collaborative Research Center TR27, the excellence cluster ``Origin and Structure of the Universe'', and the Maier-Leibnitz-Laboratorium Garching in Germany;
the Russian Academy of Science, the Kurchatov Institute and RFBR (the Russian Foundation for Basic Research);
the Brazilian Ministry of Science, Technology and Innovation (MCTI), the Financiadora de Estudos e Projetos (FINEP), the Conselho Nacional de Desenvolvimento Cient\'ifico e Tecnol\'ogico (CNPq), the S\~ao Paulo Research Foundation (FAPESP), and the Brazilian Network for High Energy Physics (RENAFAE) in Brazil.



\end{document}